\documentclass[11pt,fleqn,twoside]{article}

\usepackage[german,english]{babel}
\usepackage{isreport}

\usepackage{times}
\usepackage{alltt}
\usepackage{url}
\usepackage{xspace}
\usepackage{latexsym}

\newcommand{\dlv}{\texttt{\small{DLV}}\xspace}

\foreignreport

\title{Computing Preferred Answer Sets by Meta-Interpretation in Answer Set Programming}

\authornames{
Thomas Eiter
\and
Wolfgang Faber
\and 
Nicola Leone
\and 
Gerald Pfeifer
}

\author{
Thomas Eiter%
\affiliation{
Institut f{\"u}r Informationssysteme, Abteilung Wissensbasierte Systeme,
Technische Universit{\"a}t Wien,\protect\newline
Favoritenstra\ss{}e\ 9-11, A-1040 Vienna, Austria.
\mbox{E-mail: $\{$eiter, faber$\}$@kr.tuwien.ac.at.}}, 
~~Wolfgang Faber%
\footnotemark[1], 
Nicola Leone%
\affiliation{Department of Mathematics, University of Calabria,
87030 Rende (CS), Italy.
\mbox{E-mail: leone@unical.it.}}, 
Gerald Pfeifer%
\affiliation{
Institut f{\"u}r Informationssysteme, Abteilung Datenbanken und AI,
Technische Universit{\"a}t Wien,\protect\newline
Favoritenstra\ss{}e\ 9-11, A-1040 Vienna, Austria.
\mbox{E-mail: pfeifer@dbai.tuwien.ac.at.}}}

\published{Preliminary results of this paper appeared in Alessandro Provetti and Tran Cao Son, editors, {\em
Proceedings AAAI 2001 Spring Symposium on Answer Set Programming:
Towards Efficient and Scalable Knowledge Representation and
Reasoning}, Stanford CA, March 2001, AAAI Press, ISBN 1-57735-129-0.}

\acknowledgement{The authors would like to thank our colleagues and the participants of
the AAAI 2001 Spring Symposium on Answer Set Programming for their
comments on this work. This work was supported by the Austrian Science
Fund (FWF) under grants P13871-INF, %
P14781-INF, and Z29-INF.}

\abstract{
Most recently, Answer Set Programming (ASP) is attracting interest as
a new paradigm for problem solving. An important aspect which needs to
be supported is the handling of preferences between rules, for which
several approaches have been presented. In this paper, we consider the
problem of implementing preference handling approaches by means of
meta-interpreters in Answer Set Programming. In particular, we
consider the preferred answer set approaches by Brewka and Eiter, by
Delgrande, Schaub and Tompits, and by Wang, Zhou and Lin.  We present
suitable meta-interpreters for these semantics using \dlv, which is
an efficient engine for ASP. Moreover, we also present a
meta-interpreter for the weakly preferred answer set approach by
Brewka and Eiter, which uses the weak constraint feature of \dlv as a
tool for expressing and solving an underlying optimization problem.
We also consider advanced meta-interpreters, which make use of
graph-based characterizations and often allow for more efficient
computations. Our approach shows the suitability of ASP in general and
of \dlv in particular for fast prototyping. This can be fruitfully
exploited for experimenting with new languages and
knowledge-representation formalisms.
}

\infsysrr{1843-02-01}
\date{January 2002}

\newenvironment{proof}[1][Proof]
  {\noindent\emph{#1}\\}
  {\hspace*{1em}\ensuremath{\Box}}

\newcommand{\nop}[1]{}

\newcommand{\tuple}[1]{(#1)} 
\newcommand{\dr}[2]{^{#1}\!{#2}} 
\newcommand{\R}{\ensuremath{r}}

\newcommand{\AS}{{\cal AS}}
\newcommand{\OAS}{{\cal OAS}}
\newcommand{\FP}{{\cal FP}}
\newcommand{\BPAS}{{\cal BP\!AS}}
\newcommand{\WPAS}{{\cal WP\!AS}}
\newcommand{\DPAS}{{\cal DP\!AS}}
\newcommand{\weakPAS}{w{\cal P\!AS}}

\newcommand{\ms}{\medskip}

\newtheorem{example}{Example}

\usepackage{supertabular}

{\end{supertabular}%
} 

\newenvironment{tprogram}
{\begin{tabular}{r@{\,}c@{\,}l}%
}%
{\end{tabular}%
} 

\newcommand{\fact}   [1]{{\it \ensuremath{#1}}.}
\newcommand{\clause} [2]{{\it \ensuremath{#1}} & \derives  & {\it \ensuremath{#2}.}}
\newcommand{\wclause}[2]{ & \wderives & {\it \ensuremath{#1}.}\ [#2]}
\newcommand{\clauseB}[2]{{\it \ensuremath{#1}} & \derives & {\it \ensuremath{#2}}}
\newcommand{\clauseM}[1]{& &{\it \ensuremath{#1}}}
\newcommand{\clauseE}[1]{& &{\it \ensuremath{#1}.}}

\newcommand{\tneg}{\ensuremath{\neg}}
\newcommand{\naf}{\ensuremath{\mathtt{not}}\ }
\newcommand{\Or}{\ensuremath{\ \mathtt{v}\ }}
\newcommand{\derives}{\mbox{\texttt{:-}}\xspace}
\newcommand{\wderives}{\ensuremath{:\sim}}

\newcommand{\Ground}[1]{\ensuremath{Ground(#1)}}

\newcommand{\GR}{\ensuremath{\Ground{\R}}}
\newcommand{\HR}{\ensuremath{H(\R)}}
\newcommand{\BR}{\ensuremath{B(\R)}}
\newcommand{\BpR}{\ensuremath{B^+(\R)}}
\newcommand{\BnR}{\ensuremath{B^-(\R)}}

\renewcommand{\P}{\ensuremath{P}\xspace}
\newcommand{\WCP}{\ensuremath{WC(\P)}}
\newcommand{\GP}{\ensuremath{\Ground{\P}}}
\newcommand{\BP}{\ensuremath{B_{\P}}}
\newcommand{\UP}{\ensuremath{U_{\P}}}

\newcommand{\p}{\ensuremath{{\cal P}}\xspace}

\newcommand{\Cb}[1]{\ensuremath{C^B_{#1}}}
\newcommand{\Cw}[1]{\ensuremath{C^W_{#1}}}
\newcommand{\Cd}[1]{\ensuremath{C^D_{#1}}}

\newcommand{\PIa}{\ensuremath{P_{I_a}}\xspace}
\newcommand{\PIb}{\ensuremath{P_{I_B}}\xspace}
\newcommand{\PIw}{\ensuremath{P_{I_W}}\xspace}
\newcommand{\PId}{\ensuremath{P_{I_D}}\xspace}
\newcommand{\PIweak}{\ensuremath{P_{I_{\mathrm{weak}}}}\xspace}
\newcommand{\PIg}{\ensuremath{P_{I_g}}\xspace}

\newcommand{\wmax}[1]{\ensuremath{w_{max}^{#1}}}
\newcommand{\lmax}[1]{\ensuremath{l_{max}^{#1}}}

\newtheorem{definition}{Definition}
\newtheorem{theorem}{Theorem}
\newtheorem{proposition}{Proposition}

\newcommand{\code}[1]{\ensuremath{#1}}

\begin{document}

\maketitle

\section{Introduction}
\label{sec:intro}

Handling preference information plays an important role in
applications of knowledge representation and reasoning. In the context
of logic programs and related formalisms, numerous approaches for
adding preference information have been proposed, including
\cite{baad-holl-95,brew-94,brew-96,bucc-etal-99a,delg-scha-00,delg-etal-00a,gelf-son-97,mare-trus-93a,rint-98,saka-inou-00,wang-etal-2000,zhan-foo-97a}
to mention some of them. These approaches have been designed for
purposes such as capturing specificity or normative preference; see
e.g.\ \cite{brew-eite-99,delg-scha-00,saka-inou-00} for reviews and
comparisons.

The following example is a classical situation for the use of
preference information.

\begin{example}[bird \& penguin] Consider the following logic program:

\begin{tprogram}
(1) \hfill \fact{peng} \\
(2) \hfill \fact{bird} \\
(3)\quad \hfill \clause{\neg flies}{\naf flies,\, peng} \\
(4) \hfill \clause{flies}{\naf \neg flies,\, bird} 
\end{tprogram}

This program has two answers sets: $A_1 = \{peng,$ $bird,$ $\neg
flies\}$ and $A_2 = \{peng,$ $bird,$ $flies\}$. Assume that rule $(i)$
has higher priority than $(j)$ iff $i<j$ (i.e., rule (1) has the
highest priority and rule (4) the lowest). Then, $A_2$ is no longer
intuitive, as $flies$ is concluded from (4), while
(3) has higher priority than (4), and thus $\neg flies$ should
be concluded.
\end{example}

However, even if this example is very simple, the various preference
semantics arrive at different results. Furthermore, semantics which
coincide on this example may well yield different results on other
examples. Since evaluating a semantics on a number of benchmark
examples, each of which possibly involving several rules, quickly becomes
a tedious task, one would like to have a (quick)
implementation of a semantics at hand, such that experimentation with
it can be done using computer support. Exploring a (large) number of
examples, which helps in assessing the behavior of a semantics, may
thus be performed in significantly shorter time,  and less error
prone, than by manual evaluation.  

In this paper, we address this issue and explore the implementation of
preference semantics for logic programs by the use of a powerful
technique based on Answer Set Programming (ASP), which can be seen as
a sort of {\em meta-programming in ASP}. In this technique, a given
logic program $\p$ with preferences is encoded by a suitable set of
facts $F(\p)$, which are added to a ``meta-program'' $P_I$, such that
the intended answer sets of $\p$ are
determined by the answer sets of the logic program $P_I \cup
F(\p)$. The salient feature is that this $P_I$ is universal, i.e., it
is the same for all input programs $\p$. We recall that
meta-interpretation is well-established in Prolog-style logic
programming, and is not completely new in ASP; a similar technique has
been applied previously in \cite{gelf-son-97} for defining the
semantics of logic programming with defeasible rules
(cf.\ Section~\ref{sec:rel-work}).

We focus in this paper on three similar, yet different semantics for
prioritized logic programs, namely the preference semantics by Brewka
and Eiter \cite{brew-eite-99}, Wang, Zhou and Lin
\cite{wang-etal-2000}, and Delgrande, Schaub and Tompits
\cite{delg-etal-00a}, which we refer to as B-preferred, W-preferred,
and D-preferred answer set semantics, respectively. We present ASP
meta-programs \PIb, \PIw, and \PId such that the answer sets of $\PIb
\cup F(\p)$, $\PIw \cup F(\p)$, and $\PId \cup F(\p)$ correspond
(modulo a simple projection function) precisely to the B-, W-, and
D-preferred answer sets of $\p$.  This way, by running $F(\p)$
together with the corresponding meta-program on the \dlv system
\cite{fabe-etal-99b,dlv-web}, we compute the preferred answer sets of
$\p$ in a simple and elegant way.
For B-preferred answer sets we also provide an alternate meta-program
\PIg, which implements a graph-based algorithm that deterministically
checks preferredness of an answer set and is more efficient in general.
Note that, by suitable adaptions of the meta-programs, other ASP
engines such as Smodels \cite{niem-etal-2000a} can be used as well.

B-preferred answer set semantics refines previous approaches for
adding preferences to default rules in \cite{brew-94,brew-96}. It is
defined for answer sets of extended logic programs \cite{gelf-lifs-91}
and is generalized to Reiter's default logic in
\cite{brew-eite-98b}. An important aspect of this approach is that the
definition of preferred answer sets was guided by two general
principles which, as argued, a preference semantics should satisfy. As
shown in \cite{brew-eite-99}, B-preferred answer sets satisfy these
principles, while almost all other semantics do not. W- and
D-preferred answer set semantics increasingly strengthen B-preferred
answer set semantics \cite{scha-wang-2001}. 

Since, in general, programs having an answer set may lack a preferred
answer set, also a relaxed notion of weakly preferred answer sets was
defined in \cite{brew-eite-99}.  For implementing that semantics, we
provide a meta-program $\PIweak$, which takes advantage of the {\em weak
constraints} feature \cite{bucc-etal-97a} of \dlv.

The work reported here is important in several respects:

\begin{itemize}
\item
We put forward the use of ASP for experimenting
new semantics by means of a meta-interpretation technique.  The
declarativity of logic programs (LPs) provides a new, elegant way of writing
meta-interpreters, which is very different from Prolog-style
meta-interpretation. Thanks to the high expressiveness of (disjunctive) LPs
and \dlv's weak constraints,
meta-inter\-preters can be written in a simple and declarative fashion.

\item
The description of the meta-programs for implementing the
various preference semantics also has a didactic value: it is a good
example for the way how meta-interpreters can be built using ASP. In
particular, we also develop a core meta-program for plain extended
logic programs under answer set semantics, which may be used as a
building block in the construction of other meta-programs.

\item
Furthermore, the meta-interpreters provided are
relevant per se, since they provide an actual implementation of
preferred and weakly preferred answer sets and allow for easy
experimentation of these semantics in practice.  To our knowledge,
this is the first implementation of weakly preferred answer sets. An
implementation of preferred answer sets (also on top of \dlv) has been
reported in \cite{delg-etal-00}, by mapping programs into the
framework of compiled preferences \cite{delg-scha-97}. Our
implementation, as will be seen, is an immediate translation of the
definition of preferred answer sets into \dlv code.
Weak constraints make
the encoding of weakly preferred answer sets extremely simple and
elegant, while that task would have been much more cumbersome otherwise.
\end{itemize}

In summary, the experience reported in this paper confirms the power
of ASP.  It suggests the use of the \dlv system as a high-level
abstract machine to be employed also as a powerful tool for
experimenting with new semantics and novel KR languages.

It is worthwhile noting that the meta-interpretation approach
presented here does not aim at efficiency;
rather, this approach fosters {\em simple and very fast prototyping},
which is useful e.g.\ in the process of designing and experimenting
new languages.

The structure of the remainder of this paper is as follows: In the
next section, we provide preliminaries of extended logic
programming and answer set semantics. We then develop in
Section~\ref{sec:meta-asp} a basic meta-interpreter program for extended logic
programs under answer set semantics. After that, we consider in
Section~\ref{sec:pas} meta-interpreter programs for B-preferred, W-preferred,
and D-preferred answer answer sets. The subsequent
Section~\ref{sec:wpas} is devoted to the refinement of preferred
answer sets to weakly preferred answer sets.
A discussion of related work is provided in
Section~\ref{sec:rel-work}. The final Section~\ref{sec:conclusion}
summarizes our results and draws some conclusions.

\section{Preliminaries: Logic Programs}
\label{sec:prelim}

\paragraph{Syntax.} Logic Programs (LPs) use a function-free first-order
language. As
for terms, strings starting with uppercase (resp., lowercase) letters
denote variables (resp., constants). A ({\em positive} resp.\ {\em
negative}) {\em classical literal} $l$ is either an atom $a$ or a
negated atom $\tneg a$, respectively; its {\em complementary} literal,
denoted $\tneg l$, is $\tneg a$ and $a$, respectively. A ({\em
positive} resp.,\ {\em negative}) {\em negation as failure (NAF)
literal} $\ell$ is of the form $l$ or $\naf\ l$, where $l$ is a
classical literal. Unless stated otherwise, by {\em literal} we mean a
classical literal.

A {\em rule} \R{} is a formula
\begin{equation}
a_1\ \Or\ \cdots\ \Or\ a_n\ \derives\ 
        b_1,\cdots, b_k,\ 
        \naf\ b_{k+1},\cdots,\ \naf\ b_m.
\label{rule}        
\end{equation}
where all $a_i$ and $b_j$ are classical literals and $n\geq 0,$ $m\geq
k\geq 0$.  The part to the left of ``\derives{}'' is the {\em head},
the part to the right is the {\em body} of \R{}; we omit
``\derives{}'' if \mbox{$m=0$}.  We let
$H(r)=\{a_1,\ldots$, $a_n \}$ be the set of head literals and
\mbox{$\BR{}=\BpR \cup \BnR$} the set of body literals, where $\BpR=$
$\{b_1,$\ldots, $b_k\}$ and $\BnR=\{b_{k+1},$ \ldots, $b_m \}$ are the
sets of positive and negative body literals, respectively.
An {\em integrity constraint} is a rule where $n=0$.

A {\em datalog program (LP)} {\P} is a finite set of rules. We call {\P}
{\em positive}, if
{\P} is \naf-free (i.e.\ $\forall r \in \P: B^-(r) = \emptyset$);
and {\em normal}, if {\P} is $\Or$-free (i.e.\ $\forall r \in
\P: |H(r)| \leq 1$).

A \textit{weak constraint} \R{} is an expression of the form 
\[
\wderives b_1,\cdots, b_k, \naf\ b_{k+1},\cdots,\ \naf\ b_m.\ [w:l]
\]
where every $b_i$ is a literal and $l \geq 1$ is
the \textit{priority level} and $w\geq 1$ the \textit{weight} among
the level. Both $l$ and $w$ are integers and set to 1 if omitted. The
sets $\BR{}$, $\BpR{}$, and $\BnR{}$ are defined by viewing \R{} as
an integrity constraint.
\WCP{} denotes the set of weak constraints in {\P}.

As usual, a term (atom, rule,...) is {\em ground}, if no variables
appear in it.

\paragraph{Semantics.} Answer sets for LPs with weak constraints are
defined by extending consistent answer sets for LPs as introduced in
\cite{gelf-lifs-91,lifs-96}.  We proceed in three steps: we first
define answer sets (1) of ground positive programs, then (2)
of arbitrary ground programs, and (3) finally (optimal) answer
sets of ground programs with weak constraints. As usual, the (optimal)
answer sets of a non-ground program ${\P}$ are those of its ground
instantiation \GP, defined below.

For any program \P, let \UP{} be its Herbrand universe and \BP{} be
the set of all classical ground literals from predicate symbols in
\P over the constants of \UP{}; if no constant appears in \P, an
arbitrary constant is added to \UP. For any clause \R{}, let
\GR{} denote the set of its ground instances. Then, $\GP{} =
\bigcup_{\R \in \P} \GR$. Note that ${\P}$ is ground iff \( {\P} =
\GP{} \). An interpretation is any set $I \subseteq \BP{}$ of ground
literals. It is consistent, if $I \cap \{ \tneg l \mid l\in I\} =
\emptyset$.

In what follows, let ${\P}$ be a ground program.

(1) A consistent%
\footnote{We only consider {\em consistent
answer sets}, while in \cite{lifs-96} also the (inconsistent) set $\BP{}$
may be an answer set.}
interpretation $I \subseteq \BP$ is called {\em closed} under a 
positive program {\P}, if  $\BR
\subseteq I$ implies $\HR
\cap I \neq \emptyset$ for every $\R \in \P$. A set $X \subseteq \BP$ is an {\em
answer set} for {\P} if it is a minimal set (wrt.\ set inclusion) closed under {\P}.

(2) Let $\P^I$ be the {\em Gelfond-Lifschitz reduct} of a
program {\P} w.r.t.\ $I \subseteq \BP$, i.e., the program  obtained
from {\P} by deleting 

\begin{itemize}
\item all rules $\R \in \P$ such that $\BnR \cap I \neq \emptyset$,
and 
\item all negative body literals from the remaining rules.
\end{itemize}

Then, $I \subseteq \BP$ is an answer set of {\P} iff $I$ is an answer
set of $\P^I$.  By $\AS(\P)$ we denote the set of all answer sets of
${\P}$.

\begin{example}\label{exa:artificial}
The program
$$
\begin{tprogram}
\fact{a \Or b} \\
\fact{b \Or c} \\
\clause{d \Or \tneg d}{a, c}
\end{tprogram}
$$
\noindent has three answer sets: $\{a, c, d\}$, $\{a, c, \tneg d\}$, and
$\{b\}$.
\end{example}

(3) Given a program ${\P}$ with weak constraints, we are interested in the answer sets
of the part without weak constraints which minimize the sum of weights of
the violated constraints in the
highest priority level, and among them those which minimize the sum of
weights of the violated constraints in the next lower level,
etc. This is expressed by an objective function for \P and an
answer set $A$:
\[
\begin{array}{l@{\ }c@{\ }l}
f_\P(1) & = & 1\\
f_\P(n) & = & f_\P(n-1) \cdot |\WCP| \cdot \wmax{\P} + 1,\quad n > 1\\
H_A^\P  & = & \sum_{i=1}^{\lmax{\P}} ( f_\P(i) \cdot \sum_{N \in N^{A,\P}_i} w_N)
\end{array}
\]
where \wmax{\P} and \lmax{\P} denote the maximum weight and maximum
level of a weak constraint in {\P}, respectively; $N^{A,\P}_i$ denotes
the set of weak constraints in level $i$ which are violated by $A$,
and $w_N$ denotes the weight of the weak constraint $N$. Note that $|\WCP|
\cdot \wmax{\P} + 1$ is greater than the sum of all weights in the
program, and therefore guaranteed to be greater than any sum of
weights of a single level. If weights in level $i$ are multiplied by
$f_\P(i)$, it is sufficient to calculate the sum of these updated
weights, such that the updated weight of a violated constraint of a
greater level is always greater than any sum of updated weights of
violated constraints of lower levels.

Then, $A$ is an {\em (optimal) answer set} of \P, if $A \in
\AS(\P\setminus\WCP)$ and $H_A^\P$ is minimal over
$\AS(\P\setminus\WCP)$. Let $\OAS(\P)$ denote the set of all optimal
answer sets.

\begin{example} \label{exa:artificialweak}
Let us enhance the program from Example~\ref{exa:artificial} by the 
following three weak constraints:

\begin{tprogram}
\wclause{a, c}{2:1} \\
\wclause{\tneg d}{1:1} \\
\wclause{b}{3:1}
\end{tprogram}

The resulting program $\P_{\ref{exa:artificialweak}}$ has the single optimal
answer set $A_{\ref{exa:artificialweak}} = \{a, c, d\}$ with weight 2 in
level 1.
\end{example}

\section{Meta-Interpreting Answer Set Programs}\label{sec:meta-asp}

In this section we show how a normal propositional answer set program can be
encoded for and interpreted by a generic meta-interpreter based on the
following idea:

We provide a representation $F(\P)$ of an arbitrary normal propositional
program $\P$\ \footnote{We assume that integrity constraints \code{\derives\ C.} are written as equivalent rules \code{bad \derives C, \naf bad.} where \code{bad} is a predicate not occurring otherwise in \P.} as a set of facts and combine these facts with a generic
answer set program \PIa such that
$\AS(\P) = \{\pi(A) \mid A \in \AS(F(\P) \cup \PIa) \}$,
where $\pi$ is a simple projection function.

\subsection{Representing An Answer Set Program} 

First we translate the propositional answer set program \P into a set of
facts $F(\P)$ as follows:

\begin{enumerate}
\item For each rule%
$$
\begin{tprogram}
\hfill\clause{c}{a_1, \ldots, a_m, \naf b_1, \ldots, \naf b_n}
\end{tprogram}
$$
of the program \P, $F(\P)$ contains the following facts:
$$
\begin{array}{l@{\ }l}
rule(r).\ head(c,r). & pbl(a_1,r).\ \ldots\ pbl(a_m,r).\\
                     & nbl(b_1,r).\ \ldots\ nbl(b_n,r).
\end{array}
$$
where $r$ is a unique rule identifier.

\item For each pair of complementary literals $\ell,\tneg \ell$ occurring
in the program \P we explicitly add a fact $compl(\ell,\tneg \ell).$
\end{enumerate}

\begin{example}\label{exa:penguincode}
The program of the bird \& penguin example is represented
by the following facts representing the rules:

$$
\begin{tabular}{ll}
\fact{rule(r1)} & \fact{head(peng,r1)} \\
\fact{rule(r2)} & \fact{head(bird,r2)} \\
\fact{rule(r3)} & \fact{head(neg\_flies,r3)} \\
                & \fact{pbl(peng,r3)} \ \fact{nbl(flies,r3)} \\
\fact{rule(r4)} & \fact{head(flies,r4)} \\
                & \fact{pbl(bird,r4)} \ \fact{nbl(neg\_flies,r4)} \\
\end{tabular}
$$

and the following facts representing complementary literals:

$$
\begin{tabular}{l}
\fact{compl(flies,neg\_flies)}
\end{tabular}
$$

\end{example}

\subsection{Basic Meta-interpreter program} 

Several meta-interpreters that we will encounter in the following sections
consist of two parts: a meta-interpreter program \PIa for representing an
answer set,
and another one for checking preferredness. In this section we will provide
the first part which is common to many meta-interpreters shown in this
paper.

\paragraph{Representing an answer set.} We define a predicate $in\_AS(.)$
which is true for the literals in an answer set of \P. A literal is
in an answer set if it occurs in the head of a rule whose positive
body is definitely true and whose negative body is not false.
$$
\begin{array}{l@{\ }l}
in\_AS(X)\ \derives\ head(X,R), & pos\_body\_true(R),\\
                              & \naf neg\_body\_false(R).
\end{array}
$$
The positive part of a body is true, if all of its literals are in the
answer set. Unfortunately we cannot encode such a universal
quantification in one rule. We can identify a simple case: If there
are no positive body literals, the body is trivially true.
$$
\begin{tprogram}
  \clause{pos\_body\_exists(R)}{pbl(X,R)}\\
  \clause{pos\_body\_true(R)}{rule(R), \naf pos\_body\_exists(R)}
\end{tprogram}
$$
However, if positive body literals exist, we will proceed iteratively. To
this end, we use \dlv's built-in total order on constants for defining  a
successor relation on the positive body literals of each rule, and to
identify the first and last literal, respectively, of a positive
rule body in this total order. Technically, it is
sufficient to define auxiliary relations as follows.
$$
\begin{tprogram}
  \clause{pbl\_inbetween(X,Y,R)}{pbl(X,R), pbl(Y,R), pbl(Z,R), X < Z, Z < Y}\\
  \clause{pbl\_notlast(X,R)}{pbl(X,R), pbl(Y,R), X < Y}\\
  \clause{pbl\_notfirst(X,R)}{pbl(X,R), pbl(Y,R), Y < X}
\end{tprogram}
$$
This information can be used to define the notion of the positive body
being true up to (w.r.t.\ the built-in order) some positive body
literal. If the positive body is true up to the last literal, the
whole positive body is true.
$$
\begin{tprogram}
  \clause{pos\_body\_true\_upto(R,X)}{pbl(X,R), \naf pbl\_notfirst(X,R), in\_AS(X)}\\
  \clauseB{pos\_body\_true\_upto(R,X)}{pos\_body\_true\_upto(R,Y), pbl(X,R), in\_AS(X),}\\
  \clauseE{Y < X, \naf pbl\_inbetween(Y,X,R)}\\
  \clause{pos\_body\_true(R)}{pos\_body\_true\_upto(R,X), \naf pbl\_notlast(X,R)}
\end{tprogram}
$$
The negative part of a body is false, if one of
its literals is in the answer set.
$$
\begin{tprogram}
  \clause{neg\_body\_false(Y)}{nbl(X,Y), in\_AS(X)} 
\end{tprogram}
$$
Each answer set needs to be consistent; we thus add an integrity
constraint which rejects answer sets containing complementary literals.
$$
\begin{tprogram}
\clause{}{compl(X,Y), in\_AS(X), in\_AS(Y)}
\end{tprogram}
$$
The rules described above (referred to as \PIa in the sequel) are all we need for representing answer
sets. Each answer set of $\PIa \cup F(\P)$ represents an
answer set of \P. Let $\pi$ be defined by $\pi(A) = \{ \ell \mid
in\_AS(\ell) \in A\}$. Then we can state the following:

\begin{theorem}
\label{theo:1}
  Let \P be a normal propositional program. Then, (i)
  if $A \in \AS(\PIa \cup F(\p))$ then $\pi(A) \in \AS(\P)$, and (ii)
  for each $A \in \AS(\P)$, there exists a single $A' \in \AS(\PIa \cup
  F(\P))$ such that $\pi(A') = A$.
\end{theorem}

\begin{proof}
\noindent (i)\ $\pi(A)$ must be a consistent set of literals from
\P, since for each $\ell$ s.t.\ $in\_AS(\ell) \in A$, $head(\ell,r)$
must hold for some rule $r$, which, by construction of $F(\P)$, only
holds for $\ell \in \BP$, and since the constraint $\derives\
compl(\ell,\tneg \ell), in\_AS(\ell), in\_AS(\tneg \ell).$ must be
be satisfied by $A$ (again by construction of $F(\P)$) $\{in\_AS(\ell),
in\_AS(\tneg \ell)\} \not\subseteq A$ and hence $\{\ell,\tneg \ell\}
\not\subseteq \pi(A)$ for all $\ell \in \BP$.

Thus, to show that $\pi(A) \in \AS(\P)$, it suffices to show that
($\alpha$) $\pi(A)$ is closed under $\P^{\pi(A)}$, and that ($\beta$)
$\pi(A)\subseteq T_{\P^{\pi(A)}}^\infty$ must hold, where
$T_{\P^{\pi(A)}}$ is the standard $T_P$ operator for
$P=\P^{\pi(A)}$. Let, for convenience, denote $Q = \Ground{\PIa \cup
F(\P)}$.

As for ($\alpha$), we show that if $r\in \P^{\pi(A)}$ such that
$B(r)\subseteq \pi(A)$, then there is a rule $h_r\in Q$ such that
$H(h_r)= \{in\_AS(h)\}$ where $H(r) = \{h\}$ and $h_r$ is applied in
$A$, i.e., $A\models B(h_r)$ and $in\_AS(h)\in A$.  Let $r$ stem from
a rule $r'\in \P$. Then, let $h_r$ be the (unique) rule in $Q$ such
that $H(h_r) = \{in\_AS(h)\}$ (note that $H(r)=H(r')=\{h\}$) and
$head(h,r') \in B(h_r)$. Since $B(r) = B^+(r') \subseteq \pi(A)$, we
have for each $\ell \in B(r)$ that $in\_AS(\ell) \in A$ and, by
construction of $F(\P)$, that $pbl(\ell,r')\in A$. Since $<$ induces a
linear ordering of $B^+(r')$, it follows by an inductive argument
along it that $pos\_body\_true\_upto(r',\ell) \in A$ holds for each
$\ell \in B^+(r')$ and that $pos\_body\_true(r')\in A$. Furthermore,
since $B^-(r')\cap \pi(A)=\emptyset$, it holds that $A\not\models
B(n_r)$ for each rule $n_r\in Q$ such that $H(n_r) =
neg\_body\_false(r')$. Therefore, $neg\_body\_false(r')\notin
A$. Since $head(h,r') \in A$ by construction of $F(\P)$, it follows
that $A\models B(h_r)$. Thus, $h_r$ is applied in $A$, and hence
$H(h_r) = in\_AS(h) \in A$. This proves ($\alpha$).

As for ($\beta$), it suffices to show that if $in\_AS(\ell) \in
T^i_{Q^A}\setminus T^{i-1}_{Q^A}$, i.e., $in\_AS(\ell)$ is added to
$A$ in the $i$-th step of the least fixpoint iteration for $T_{Q^A}$,
$i\geq 1$, then $\ell \in T^\infty_{\P^{\pi(A)}}$ holds. Addition of
$in\_AS(\ell)$ implies that $\ell=H(r)$ for some $r\in \P$ such that
$neg\_body\_false(r)\notin A$ and $pos\_body\_true(r)\in A$. This
implies $in\_AS(\ell') \in T^{i-1}_{Q^A}$ for each $\ell' \in B^+(r)$
and $in\_AS(\ell')\notin A$ for each $\ell' \in B^-(r)$. By an
inductive argument, we obtain $B^+(r)\subseteq
T^\infty_{\P^{\pi(A)}}$; therefore, $\ell \in T^\infty_{\P^{\pi(A)}}$
holds. This proves ($\beta$).

\medskip
  
\noindent (ii)\ For any $A \in \AS(\P)$, let $A'$ be defined as
follows ($<$ is the total order on constants defined in \dlv):

\begin{tabbing}
$A'$ = \= $\{in\_AS(x) \mid x \in A \}\ \cup$ \+\\
$\{pos\_body\_true\_upto(r,x) \mid x\!\in\!\BpR \land x\!\in\!A \land \forall y\!\in\!\BpR: y\!<\!x \rightarrow y\!\in\!A\}\, \cup$\\
$\{pos\_body\_true(r) \mid \forall x \in \BpR: x \in A \}\ \cup$\\
$\{neg\_body\_false(r) \mid \BnR \cap A \neq \emptyset\}\ \cup$\\
$\{pos\_body\_exists(r) \mid \BpR \neq \emptyset \}\ \cup$\\
$\{pbl\_notfirst(x,r) \mid x \in \BpR \land \exists y \in \BpR: y < x \}\ \cup$\\
$\{pbl\_notlast(x,r) \mid x \in \BpR \land \exists y \in \BpR: x < y \}\ \cup$\\
$\{pbl\_inbetween(x,y,r) \mid x,y \in \BpR \land ( \exists z \in \BpR: x < z \land z < y )\}\ \cup$\\
$F(P)$.
\end{tabbing}

Observe that the set of literals defined by the last five lines (call
it $A_{stat}$) do not depend on $A$ since they have to occur in all
answer sets of $\PIa \cup F(\P)$. The definitions of $A_{stat}$
directly reflect the corresponding rule structure in $\PIa$ and $F(\P)$. Since the
inclusion of literals $in\_AS(x)$ into $A'$ is determined by the condition
$\pi(A')=A$, it is easy to see that all rules defining
\code{neg\_body\_false} are satisfied; in the case of
\code{pos\_body\_true\_upto} and \code{pos\_body\_true}, rule satisfaction can
be seen by a constructive argument along the order $<$. 

To see that $A'$ is minimal and the only answer set of $\PIa \cup
F(\P)$ s.t.\ $\pi(A')=A$, an argument similar to the one in ($\beta$)
of the proof for (i) can be applied: If another answer set $A''$
exists s.t.\ $\pi(A'')=A$, each $in\_AS(x)$ s.t.\ $x \in A$ must be
added in some stage of $T^i_{Q^{A'}}$, and in some stage
$T^j_{Q^{A''}}$ of the standard least fixpoint operator. But then, by
an inductive argument, $T^\infty_{Q^{A'}} = T^\infty_{Q^{A''}}$ must
hold.
\end{proof}

The meta-interpreter program \PIa has the benign property that a standard
class of programs, namely the class of stratified programs, which are
easy to evaluate, is also interpreted efficiently through it. Recall that
a normal propositional program \P is {\em stratified}, if there is a function $\lambda$
which associates with each atom $a$ in \P an integer
$\lambda(a)\geq 0$, such that each rule $r \in \P$ with $H(r) = \{h\}$
satisfies $\lambda(h)\geq
\lambda(\ell)$ for each $\ell\in B^+(r)$ and $\lambda(h)>
\lambda(\ell)$ for each $\ell\in B^-(r)$.
Denote for any program \P by $\P_{ir}$ the
set of rules in $\Ground{\PIa \cup F(\P)}$ whose representation
literals (i.e., literals over $rule$, $head$, $bpl$, $nbl$) are
satisfied by $F(\P)$. Then we have:

\begin{proposition}\label{prop:strat}
Let \P be a stratified normal propositional program. Then, $\P_{ir}$
is locally stratified (i.e., stratified if viewed as a propositional program).
\end{proposition}

This can be seen by constructing from a stratification mapping
$\lambda$ for \P a suitable stratification mapping $\lambda'$ for
$\P_{ir}$. Locally stratified programs are efficiently handled by
\dlv. Since, by intelligent grounding strategies, $\P_{ir}$ is
efficiently computed in \dlv, the overall evaluation of $\PIa\cup
F(\P)$ is performed efficiently by \dlv for stratified programs $\P$.

\section{Preferred Answer Sets}
\label{sec:pas}

In this section, we will first introduce the underpinnings common to all
three semantics for preferred answer sets we are considering in this
paper and then provide the individual definitions.

\subsection{Prioritized Programs}

We recall and adapt the definitions of \cite{brew-eite-99} as needed
in the current paper. Throughout the rest of this section, programs
are tacitly assumed to be propositional.

\begin{definition}[prioritized program]
\label{def:prioruleb}
A prioritized (propositional) program is a pair $\p = (P,<)$ where $P$ is a normal
logic program without constraints, and $<$ is a strict partial order
on $P$, i.e., an irreflexive ($a \not< a$, for
all $a$) and transitive relation.
\end{definition}

Informally, $r_1 < r_2$ means ``$r_1$ has higher priority than
$r_2$''.  For any $\p=(P,<)$, the answer sets of $\p$ are
those of $P$; their collection is denoted by $\AS(\p)=\AS(P)$.

\begin{definition}[full prioritization]
\label{def:full_priorization}
A full prioritization of a prioritized program $\p = (P,<)$ is any
pair $\p' = (P,<')$ where $<'$ is a total order of $P$ refining
$<$, i.e., $r_1 < r_2$ implies $r_1 <' r_2$, for all $r_1,r_2 \in
P$. The set of all full prioritizations of $\p$ is denoted by
$\FP(\p)$. We call $\p$ fully prioritized, if $\FP(\p) = \{ \p\}$.
\end{definition}

Fully prioritized programs $\p=(P,<)$ are also referred to as ordered
sets $\p=\{ r_1,\ldots,r_n\}$ of rules where $r_i < r_j$ iff $i<j$.

\subsection{B-Preferred Answer Sets}
\label{sec:b-pref}

B-preferred answer sets have been introduced in \cite{brew-eite-99} as a 
refinement of previous approaches in \cite{brew-94,brew-96}.
 
We first define B-preferred answer sets for a particular class of fully
prioritized programs. Call a program $\p$ {\em prerequisite-free}, if
$\BpR{}=\emptyset$ for every rule $\R{}\in \p$ holds. Furthermore, a
literal $\ell$ (resp., a set $X\subseteq \BP$ of literals) {\em
defeats} a rule $\R{}$ of the form (\ref{rule}), if $\ell \in \BnR{}$
(resp., $X \cap \BnR{}\neq\emptyset$). We say that a rule $r'$ defeats
a rule $r$ if $H(r')$ defeats $r$.

\begin{definition}
\label{def:s-alpha}
Let $\p = \{ r_1,\ldots,r_n\}$ be a fully prioritized and
prerequisite-free program.
For any set $S \subseteq \BP$ of literals,  the sequence $S_i\subseteq \BP$
($0 \leq i \leq n$) is defined as follows:

$$
\begin{array}{ll}
S_0= & \emptyset \\[1.2ex]
S_i= & \left\{ \begin{array}{lp{0.40\textwidth}}
  S_{i-1},            & if $(\alpha)$ $S_{i-1}$ defeats $r_i$, or \\
                      & $(\beta)$ $H(r_i) \subseteq S$ and $S$ defeats $r_i$,\\
  S_{i-1}\cup H(r_i), & \textrm{otherwise.}
               \end{array} \right.
\end{array}
$$

for all $i=1,\ldots,n$. The set $\Cb{\p}(S)$ is defined by 

$$
\Cb{\p}(S) =\left\{
\begin{array}{lp{0.30\textwidth}}
 S_n,  & if $S_n$ is consistent, \\
 \BP{} & otherwise.
\end{array}\right.
$$
\end{definition}

An answer set $A$ ($=S$) divides the rules of $P$ in
Definition~\ref{def:s-alpha} into three groups: {\em generating rules},
which are applied and contribute in constructing $A$; {\em dead
rules}, which are not applicable in $A$ but whose consequences would
not add anything new if they were applied, since they appear in $A$;
and {\em zombies}, which are the rules not applicable in $A$ whose
consequences do not belong to $A$. Only zombies have the potential to
render an answer set non-preferred. This is the case if some zombie is
not ``killed'' by a generating rule of higher priority. If $A$ is a
fixpoint of $\Cb{\p}$, then the inductive construction guarantees that
indeed all zombies are defeated by generating rules with higher
preference.

\begin{definition}[B-preferred answer set]
\label{def:full-prereqfree-pas}
Let $\p= (P,<)$ be a fully prioritized and prerequisite-free program, and let
$A \in \AS(\p)$.
Then $A$ is a B-preferred answer set of $\p$ if and only if $\Cb{\p}(A) = A$.
\end{definition}

In the case where $P$ is not prerequisite-free,
a kind of dual Gelfond-Lifschitz reduct is computed as follows.

\begin{definition}
\label{defn:dr}
Let $\p= (P,<)$ be a fully prioritized program, and let $X\subseteq
\BP$. Then $\dr{X}{\p} =
(\dr{X}{P},\dr{X}{<})$ is the fully prioritized program
such that: 
\begin{itemize}
\item  $\dr{X}{P}$ is the set of rules obtained from $P$ by deleting
\begin{enumerate}
\item every $\R\in \p$ such that $\BpR\not\subseteq X$, and
\item all positive body literals from the remaining rules.
\end{enumerate}
\item 
$\dr{X}{<}$ is inherited from $<$ by the map $f:$ $\dr{X}{P}
        \longrightarrow P$ (i.e., \mbox{$r'_1$ $\dr{X}{<}$ $ r'_2$} iff $f(r'_1) < f(r'_2)$), where $f(r')$ is the
first rule in $P$ w.r.t.\ $<$ such that $r'$ results from $r$ by Step~2.
\end{itemize}
\end{definition}

The definition of $\dr{X}{<}$ must respect possible clashes of rule
priorities, as Step~2 may produce duplicate rules in general. 

\begin{definition}[B-preferred answer set (ctd.), $\BPAS$]
\label{def:BPAS}
Let $\p = (P,<)$ be a prioritized
program and  $A \in AS(P)$. If $\p$ is fully prioritized, then $A$ is a
B-preferred answer set of $\p$ iff $A$ is a B-preferred
answer set of $\dr{A}{\p}$; otherwise, $A$ is a B-preferred answer set
of $\p$ iff $A$ is a B-preferred answer set for some $\p' \in
\FP(\p)$.
By $\BPAS(\p)$ we denote the set of all B-preferred answer sets of $\p$.
\end{definition}

\begin{example}
Reconsider the bird \& penguin example. Let us first check whether
$A_1 = \{peng,$ $bird,$ $\neg flies\}$ is a B-preferred answer set.
We determine the dual reduct
$\dr{A_1}{\p}$ which consists of the following rules: 

\medskip

\begin{tprogram}
(1) \quad\hfill \fact{peng} \\
(2) \hfill \fact{bird} \\
(3) \hfill \clause{\neg flies}{\naf flies} \\
(4) \hfill \clause{flies}{\naf \neg flies} 
\end{tprogram}

\medskip
The order $\dr{A_1}{<}$ coincides with $<$ as in
Definition~\ref{defn:dr}. Now, let us determine $A_{1,4}$ ($=S_4$), by
constructing the sequence $A_{1,i}$, for $0\leq i\leq 4$:
$A_{1,0}=\emptyset$, $A_{1,1} = \{ peng \}$, $A_{1,2} =\{ peng,
bird\}$, $A_{1,3} = \{ peng, bird, \neg flies\}$, and $A_{1,4} =
A_{1,3}$.  Thus, $A_{1,4} =$ $\{ peng,$ $bird,$ $\neg flies\} = A_1$
and $\Cb{\dr{A_1}{\p}}(A_1)=A_1$; hence, the answer set $A_1$ is
preferred.

Next consider the answer set $A_2 = \{peng,$ $bird,$ $flies\}$.
The dual reducts $\dr{A_2}{\p}$ and $\dr{A_1}{\p}$ coincide, and thus
$A_{2,4} = A_1$, which means  $\Cb{\dr{A_2}{\p}}(A_2)\neq A_2$.  Hence,
$A_2$ is not preferred, and $A_1$ is the single B-preferred answer set of $\p$.
\end{example}

The following example shows that not every prioritized program which
has an answer set has also a B-preferred one.

\begin{example}
\label{exa:nonex-prans}
Consider the following program:
\ms

\begin{tprogram}
(1) \quad\hfill \clause{c}{\naf \mbox{ $b$ }} \\ 
(2) \hfill \clause{b}{\naf \mbox{ $a$ }} 
\end{tprogram}

\ms

where $(1) < (2)$. Its single answer set is $A = \{b\}$. However,
$\Cb{\dr{A}{\p}}(A) = \{c,b\}$ and thus $A$ is not B-preferred.
\end{example}

\subsubsection{Adapting the Meta-Interpreter}

Now we can extend the meta-interpreter for answer set programs from 
Section~\ref{sec:meta-asp} to cover prioritized answer set programs.

\paragraph{Representing a prioritized program.} A prioritized program
$\p=(P,<)$ is represented by a set of facts $F(\p)$ which contains $F(P)$
plus, for each rule preference $r < r'$ that belongs to the
transitive reduction of $<$, a fact $pr(r,r')$.

\begin{example}
In the case of our bird \& penguin example, we add the following three facts:
$$
\begin{tabular}{ccc}
\fact{pr(r1,r2)}  &  \fact{pr(r2,r3)}  &  \fact{pr(r3,r4)}
\end{tabular}
$$
\end{example}

\paragraph{Checking preferredness.} According to Definition~\ref{def:BPAS}, we
have to create all fully prioritized programs $\FP(\p)$ of $\p$ to
determine its preferred answer sets. To this end, we add code to guess
a total order on the rules which refines $<$:
$$
\begin{tprogram}
\clause{pr(X,Y) \Or pr(Y,X)}{rule(X),\ rule(Y),\ X\, !\!\!= Y}\\
\clause{pr(X,Z)}{pr(X,Y),\ pr(Y,Z)}\\
\clause{}{pr(X,X)}
\end{tprogram}
$$
The rules state the axioms of totality, transitivity, and
irreflexivity of a total order. Note that it would be possible to
replace the disjunctive guessing rule by two rules involving unstratified
negation. However, the disjunctive version is more readable.

Next we build the set $\Cb{\p'}$ from Definition~\ref{def:s-alpha} where
$\p'=\dr{X}{\p}$. To this end, we do not compute the sets $S_i$ as in
the definitions -- clearly one rule can contribute at most one element
to $\Cb{\p'}$ and we represent this fact using the predicate
$lit(.,.)$.  We first observe that duplicate rules arising in the dual
reduct $\p´$ need no special care, since only the first occurrence of
a rule from $\p'$ is relevant for the value of $\Cb{\p'}$; for later
occurrences of duplicates always $S_i=S_{i-1}$ will hold.

In Definition~\ref{def:s-alpha}, a condition when $H(r_i)$ is not added is
stated, while $lit(.,.)$ represents the opposite, so we negate the
condition: $(\beta)$ is actually itself a conjunction $\gamma \land
\delta$, so the condition we are interested in is
$$
\neg(\alpha \lor
(\gamma \land \delta)) \equiv (\neg\alpha \land \neg\gamma) \lor
(\neg\alpha \land \neg\delta).
$$
We call condition $\alpha$ {\em
local defeat} (by rules of higher priority) and $\delta$ {\em global
defeat} (by the answer set).

Definition~\ref{def:s-alpha} applies only to prerequisite-free
programs, so for the general case we also have to include the
definition of the dual Gelfond-Lifschitz reduct, which amounts to
stating that only rules with a true positive body w.r.t.\ the answer set
have to be considered. The encoding is then straightforward:
$$
\begin{tprogram}
\clauseB{lit(X,Y)}{head(X,Y), pos\_body\_true(Y),}\\
&& $\naf defeat\_local(Y), \naf in\_AS(X).$\\
\clauseB{lit(X,Y)}{head(X,Y), pos\_body\_true(Y),}\\
&& $\naf defeat\_local(Y), \naf defeat\_global(Y).$
\end{tprogram}
$$
$$
\begin{tprogram}
\clause{defeat\_local(Y)}{nbl(X,Y), lit(X,Y1), pr(Y1,Y)}\\
\clause{defeat\_global(Y)}{nbl(X,Y), in\_AS(X)}
\end{tprogram}
$$
The set $\Cb{\p'}$ is the union of all literals in $lit(.,.)$:
$$
\begin{tprogram}
\clause{in\_CP(X)}{lit(X,Y)}
\end{tprogram}
$$
Finally, according to Definition~\ref{def:full-prereqfree-pas}, a
preferred answer set $A$ must satisfy $A = \Cb{\p'}(A)$, so we formulate
integrity constraints which discard answer sets violating this condition:
$$
\begin{tprogram}
\clause{}{in\_CP(X), \naf in\_AS(X)}\\
\clause{}{in\_AS(X), \naf in\_CP(X)}
\end{tprogram}
$$
\begin{figure}[tb]
{\small
\begin{alltt}
% For full prioritization: refine pr to a total ordering.
  pr(X,Y) v pr(Y,X) :- rule(X), rule(Y), X != Y. 
  pr(X,Z) :- pr(X,Y), pr(Y,Z). 
  :- pr(X,X). 

% Check dual reduct: Build sets S_i, use rule ids as indices i.
% lit(X,r) means that the literal x occurs in the set S_r.
   lit(X,Y) :- head(X,Y), pos_body_true(Y), 
               not defeat_local(Y), not in_AS(X). 
   lit(X,Y) :- head(X,Y), pos_body_true(Y), 
               not defeat_local(Y), not defeat_global(Y). 
   defeat_local(Y)  :- nbl(X,Y), lit(X,Y1), pr(Y1,Y).  
   defeat_global(Y) :- nbl(X,Y), in_AS(X). 

% Include literal into CP(.).
   in_CP(X) :- lit(X,Y).
   :- in_CP(X), not in_AS(X). 
   :- in_AS(X), not in_CP(X).  % this constraint is redundant
\end{alltt}
}
\caption{Meta-Interpreter \PIb for B-Preferred Answer Sets (without \PIa)} 
\label{fig:PIb}
\end{figure}

This completes the meta-interpreter program \PIb. A compact listing of
it (without showing \PIa explicitly) is given in
Figure~\ref{fig:PIb}. The following result states that it works
correctly.

\begin{theorem}
\label{theo:2}
Let $\p = (\P,<)$ be a propositional prioritized program. Then, (i)
if $A \in \AS(\PIb \cup F(\p))$ then $\pi(A) \in \BPAS(\p)$, and (ii)
for each $A \in \BPAS(\p)$, there exists some $A' \in \AS(\PIb \cup
F(\p))$ such that $\pi(A') = A$.
\end{theorem}

\begin{proof} Let $Q=\PIb\cup F(\p)$ and $Q_a = \PIa\cup F(\p)$. By
well-known results about splitting a logic program
\cite{lifs-turn-94}, we obtain that for each answer set $A$ of $Q$,
its restriction $A_a$ to the predicates of $Q_a$ is an answer set
$Q_a$, and that $A$ is an answer set of $(\PIb\setminus Q_a) \cup
A_a$. 

\noindent (i)\ Suppose that $A \in \AS(Q)$.  Then, $A_a$ is an answer
set of $Q_a$, and thus, by Theorem~\ref{theo:1}, $S=\pi(A_a)$
($=\pi(A)$) is an answer set of $P$. Furthermore, by the three clauses
that define and constrain the predicate \code{pr} in $\PIb$, $A$
defines a total ordering $<'$ on $P$ such that $r<'r'$ is equivalent
to \code{pr(r,r')}. Consider the dual reduct $\dr{S}{\p} =
(\dr{S}{P},\dr{S}{{<'}})$. Since $S$ is an answer set of $P$, it
remains to show that $\Cb{\dr{S}{\p}}(S) = S$ holds. Define the sets
$S_i$, $0\leq i \leq n$, along the ordering $\dr{S}{<'}$ as
follows. Denote by $r^{\dr{S}{\p}}_i \in P$ the least rule $r$ under $<'$
such that its dual reduct $\dr{S}{r}$ (i.e., $H(\dr{S}{r})=H(r)$ and
$B(\dr{S}{r})=B^-(r)$) is the $i$-th rule in $\dr{S}{\p}$
under $\dr{S}{<'}$, where $1\leq i \leq n$. Then, let $S_0=\emptyset$
and let $S_i
= \{ \ell \mid lit(\ell,r) \in A$, where
$r = r^{\dr{S}{\p}}_j$ for some $j\leq i\}$, for all $i=1,\ldots,n$.

By an inductive argument on $i=0,\ldots,n$, we obtain that the
sequence $S_0$, \ldots, $S_n$ satisfies the condition of
Definition~\ref{def:s-alpha}. Indeed, this is true for $i=0$. Suppose
it is true for $i-1$ and consider $i$. Let $r = r^{\dr{S}{\p}}_i$ where $H(r)=\{h\}$. Then,
\code{pos\_body\_true(r)} $\in A$ holds. If $h\in S_{i-1}$ holds, then
the condition in Definition~\ref{def:s-alpha} anyway holds. Thus,
suppose $h\notin S_{i-1}$ and first that $S_{i}=S_{i-1}$. Then, both
rules with head \code{lit(h,r)} in $Q$ are not applied in $A$. This
means that either ($\alpha$) \code{defeat\_local(r)} $\in A$ or
($\beta$) $\{\code{in\_AS(h)},$ $\code{defeat\_global(r)}\} \subseteq
A$ must hold. Case ($\alpha$) implies that, by the induction
hypothesis, there is some literal $\ell \in B^-(r)$ such that $\ell
\in S_{i-1}$; that is, $S_{i-1}$ defeats $r$.  Case ($\beta$) means
that $H(r) \subseteq S$ and that $S$ defeats $r$. Thus, $S_i$
satisfies Definition~\ref{def:s-alpha} in this case.

Otherwise, suppose we have $h \in S_i\setminus S_{i-1}$. Thus,
\code{lit(h,r)}$\in A$, which means that one of the
two rules with head \code{lit(h,r)} in $Q$ is applied in $A$. Thus, we
have ($\alpha$) $\code{defeat\_local(r)}\notin A$, which, by the
induction hypothesis, means that $r$ is not defeated by $S_{i-1}$, and
that ($\beta$) either \code{in\_AS(h)} $\notin A$, hence
$H(r)\not\subseteq S$, or \code{defeat\_global(r)} $\notin A$, which,
by the definition of \code{defeat\_global}, means that $S$ does not
defeat $r$. Thus, the condition of Definition~\ref{def:s-alpha} is
satisfied. This proves the claim on the sequence $S_0,\ldots,S_n$.

As easily seen, $S_n = \{ \ell \mid \exists r: \code{lit(\ell,r)} \in
A\}$. Therefore, from the rule defining \code{in\_CP} in $\PIb$, we
obtain that \code{in\_CP(\ell)} $\in A$ holds iff $\ell \in S_n$, for any
constant $\ell$. Thus, from the last two constraints of $\PIb$, we
infer that $\Cb{\dr{S}{\p}}(S) = S$ must hold; in other words,
$S=\pi(A)$ is a $B$-preferred answer set of $\p$.
\medskip

\noindent (ii)\ Suppose that $A \in \BPAS(\p)$, and let $\p'=(P,<')$ be a
full prioritization of $\p$, such that $A \in \BPAS(\p')$. Then, we
obtain an answer set $A'$  of $Q$  as follows:
\begin{itemize}
\item On the predicates defined in $\PIa\cup F(\p)$, $A'$ coincides
with the answer set of $\PIa\cup F(P)$ corresponding to $A$ as in item
(ii) of Theorem~\ref{theo:1};
\item \code{pr} is defined according to $<'$, i.e., \code{pr(r,r')}
        iff $r<'r'$; 
\item \code{lit(h,r)} is true iff $r$ is a rule from $P$ such that 
$r$ is applied in $A$;
\item \code{defeat\_local(r)} is true iff some rule $r'<'r$ exists such
 that $H(r')\cap B^-(r) \neq \emptyset$ and $r'$ is applied in
 $A$; 

\item \code{defeat\_global(r)} is true iff $A \cap B^-(r) \neq \emptyset$; 
\item \code{in\_CP(\ell)} is true iff $\ell \in A$. 
\end{itemize}
Note $A'$ satisfies the last two constraints in $\PIb$ by virtue of Theorem~\ref{theo:1}, since $A$ is
an answer set of $\p$.

By the splitting result of \cite{lifs-turn-94}, for showing that $A'$
is an answer set of $Q$, we only need to show the following. Let $A''$
be the restriction of $A'$ to the predicates in $\PIa \cup F(\p)$ and
\code{pr}; then, $A'$ is an answer set of the program $Q_1$, which
contains $A''$ plus all clauses $c$ in $Q$ which involve the
predicates \code{lit}, \code{defeat\_local}, \code{defeat\_global},
\code{in\_CP} and such that $A''$ satisfies all literals in $B(c)$ on
the predicates in $\PIa$ and \code{pr}.

As easily seen, $Q_1$ is locally stratified, and a stratification
$\lambda$ exists on the atoms occurring in $Q_1$ such that
\begin{itemize}
\item $\lambda(\code{lit(\ell',r')}) < \lambda(\code{lit(\ell,r)})$ and $\lambda(\code{defeat\_local(r')}) <
\lambda(\code{defeat\_local(r)})$, for all constants $\ell$, $\ell'$ and rules $r,r' \in P$ such that $r'<'r$;
\item $\lambda(\code{defeat\_local(r)})<\lambda(\code{lit(\ell,r)})$, for all literals $\ell$ and
rules $r\in P$;  
\item $\lambda(\code{in\_CP(\ell)})= 1+ \max \{
\lambda(\code{lit(\ell,r)}) \mid \code{head(\ell,r)}\in A'' \}$; and 
\item $\lambda(a)=0$, for all other atoms $a$ in $Q_1$. 
\end{itemize}

Then, along $\lambda$, we can verify that $Q_1$ has a stratified model
$S$ which coincides with $A'$, i.e., for $i\geq 0$,  we have for all atoms $a$
with $\lambda(a)\leq i$ that $a\in A'$ iff $a\in S$.

For $i=0$, this is immediate from the definition of $A'$. Suppose the
statement is true for $i$, and consider $i+1$. Let $a$  be an atom
such that $\lambda(a)=i+1$. Suppose first that $a\in A'$, and consider
the possible cases: 
            
\noindent$\bullet$\ If $a=\code{defeat\_local(r)}$, then some $r'<'r$
exists such that $H(r')\cap B^-(r)\neq\emptyset$ and $r'$ is applied
in $A$.  By definition, \code{lit(\ell,r')}$\in A'$ holds, and by the
induction hypothesis, \code{lit(\ell,r')}$\in S$. Thus,
\code{defeat\_local(r)}$\in S$.

\noindent$\bullet$\ Next, if $a=\code{in\_CP(\ell)}$, then by
definition of $A'$, $\ell \in A$.  Since $A$ is a $B$-preferred answer
set of $\p'$, it follows that \code{lit(\ell,r)}$\in A'$ for
some $r$ such that $\lambda(\code{lit(\ell,r)})\leq i$. Thus, by the
induction hypothesis, \code{lit(\ell,r)}$\in S$, which implies
$\code{in\_CP(\ell)}\in S$.

\noindent$\bullet$\ Finally, if $a=\code{lit(\ell,r)}$, then by
definition of $A'$, $r$ is applied in $A$.  Thus, we have by
construction \code{head(\ell,r)}$\in S$, \code{pos\_body\_true(r)}$\in
S$, \code{defeat\_global(r)}$\notin S$, and all these atoms rank lower
than \code{lit(\ell,r)}. Furthermore, \code{defeat\_local(r)}$\notin
S$ must hold; otherwise, as
$\lambda(\code{defeat\_local(r)})<\lambda(\code{lit(\ell,r)})$, by the
induction hypothesis, some rule $r'<'r$ would exist which is applied
in $A$ such that $H(r')\cap B^-(r)\neq \emptyset$, which would
contradict that $r$ is applied in $A$. This means, however, that the
second rule with head \code{lit(\ell,r)} in $Q_1$ is applied, and thus
$\code{lit(\ell,r)} \in S$.

Thus, $a\in A'$ implies $a \in S$. Conversely, suppose that $a\in S$,
and again consider the possible cases: 

\noindent$\bullet$\ If $a=\code{defeat\_local(r)}$, then it follows
that \code{lit(\ell,r')}$\in S$ for some $r'<'r$ such that $H(r')\cap
B^-(r)\neq\emptyset$. By the induction hypothesis,
\code{lit(\ell,r')}$\in A'$, which by definition of $A'$ means that
$r'$ is applied in $A$; since $H(r')\cap B^-(r)\neq\emptyset$, by
definition of $A'$ we have \code{defeat\_local(r)}$\in A'$.

\noindent$\bullet$\ Next, if $a=\code{in\_CP(\ell)}$, then
\code{lit(\ell,r)}$\in S$ exists such that
$\lambda(\code{lit(\ell,r)})\leq i$. By the induction hypothesis and
the definition of $A'$, we have that $r$ is applied in $A$. Therefore, $\ell\in A$, which by definition means
\code{in\_CP(\ell)}$\in A'$. 

\noindent$\bullet$\ Finally, if $a=\code{lit(\ell,r)}$, then
\code{head(\ell,r)}$\in A'$, \code{pos\_body\_true(r)}$\in A'$, and
\code{defeat\_local(r)}$\notin A'$, and either ($\alpha$)
\code{in\_AS(\ell)}$\notin A'$, or ($\beta$)
\code{defeat\_global(r)}$\notin A'$. In case ($\alpha$), by definition
of $A'$ we have $\ell\notin A$. This implies, however, that $A$ is not
a $B$-preferred answer set of $\p$, which is a contradiction. Thus, ($\beta$)
must apply. By the induction hypothesis, we obtain
\code{defeat\_global(r)}$\notin S$. Hence, $B^-(r)\cap A = \emptyset$,
which means that $r$ is applied in $A$; hence, \code{lit(\ell,r)}$\in
A'$ by definition.

This shows that $a\in S$ implies $a\in A'$, which concludes the
induction. We thus have shown that $S=A'$. Since $A'$ satisfies the
last two constraints of $\PIb$, it follows that $S$ is the stratified
model of $Q_1$ and $A'$ is an answer set
$Q_1$. Hence, $A'$ is an answer set of $Q$. This proves the result.
\end{proof}

\begin{example}
  For the bird \& penguin example, $\PIb \cup F(\p)$ has one answer
  set, which contains $in\_AS(peng)$, $in\_AS(bird)$, and $in\_AS(\neg
  flies)$.
\end{example}

We note that the last constraint in $\PIb$ is in fact redundant and
can be dropped; this is possible sind the fixpoint condition
$\Cb{\p}(A) = A$ in the Definition~\ref{def:full-prereqfree-pas} can be equivalently
replaced by a weaker condition. 

\begin{proposition}
Let $\p=(P,<)$ be a fully prioritized and prerequisite-free program, and let $A \in
\AS(\p)$. Then, $A\in \BPAS(\p)$ iff $\Cb{\p}(A) \subseteq A$. 
\end{proposition}

\begin{proof} It suffices to show that $\Cb{\p}(A)\subseteq
A\,\land\,\Cb{\p}(A)\neq A$ raises a contradiction. Assume the condition holds. 
Then, some $\ell \in A \setminus \Cb{\p}(A)$ must
exist, which means that a generating rule $r$ w.r.t.\ $A$ must exist
such that $H(r) = \{\ell\}$ and $A \models B(r)$. According to
Definition~\ref{def:s-alpha}, $(\alpha) \lor (\beta)$ must hold for $A_r$,
otherwise $\ell \in \Cb{\p}(A)$ would hold. Now $(\beta)$ cannot hold,
since $A$ cannot defeat $r$ because $A \models B(r)$.  Thus $(\alpha)$
must hold. This implies that some $\ell' \in A_{r-1}$ defeats $r$ such
that $\ell' \not\in A$. Sine $A_{r-1} \subseteq \Cb{\p}(A)$, it follows
$\Cb{\p}(A) \not\subseteq A$. This is a contradiction.
\end{proof}

\subsubsection{Deterministic Preferredness Checking}
\label{sec:det-checking}

The method we provided above non-deterministically generates, given a
prioritized program $\p=(P,<)$ and an answer set of $\p$, all full
prioritizations of $\p$ and tests them.

In \cite{brew-eite-99} a graph-based algorithm has been described which
checks preferredness of an answer set $A$ deterministically without
refining $<$ to a total order. In general, this method is much more
efficient. 

This approach works as follows: A labeled directed graph $G(\p,A)$ is
constructed, whose vertices are the rules $P$, and an edge leads from
$r$ to $r'$ if $r<r'$. Each vertex $r$ is labeled ``g'' if $r$ is
generating w.r.t\ $A$, ``z'' if it is a zombie, and ``i'' (for
irrelevant) otherwise. The following algorithm then performs a kind
of topological sorting for deciding whether an answer set $A$ is
preferred, and outputs a suitable full prioritization of $\p$:

\begin{description}
\item[Algorithm] FULL-ORDER 
\item[Input:] A propositional prioritized program $\p = \tuple{P,<}$, and an
  answer set $A \in \AS(P)$.
\item[Output:] A full prioritization $\p' \in \FP(\p)$ such that \mbox{$A
    \in \BPAS(\p')$} if $A \in \BPAS(\p)$; ``no'', otherwise.
\item[Method:] ~ \\
\vspace{-1\baselineskip}               
\begin{description}
\item[Step 1.] Construct the graph $G = G(\p,A)$, and initialize\ $S :=
  \emptyset$, $<' := \emptyset$. 
  
\item[Step 2.] If $G$ is empty, then output $\p' = \tuple{P,<'}$ and halt.
  
\item[Step 3.] Pick any source of $G$, i.e., a vertex $r$ with no
  incoming edge, such that either $r$ is not labeled ``z'' or $r$ is
  defeated by $S$. If no such $r$ exists, then output ``no'' and
  halt.
\item[Step 4.] If $r$ is labeled ``g'', then set $S := S \cup H(r)$.
\item[Step 5.] Remove $r$ from $G$, and continue at Step~2.
\end{description}
\end{description}

A discussion of this algorithm is given in \cite{brew-eite-99}.  Note
that it is non-deterministic in Step~3. A deterministic variant of it
can be used for merely deciding preferredness of $A$: rather than some
arbitrary source $r$, {\em all} sources $r$ satisfying the condition
are selected in Step~3 and then removed in parallel in Step~5. As
easily seen, this is feasible since removability of a source $r$ is
monotone, i.e., can not be destroyed by removing any other source $r'$
before. Thus, $A$ is a preferred answer set iff the algorithm stops
with the empty graph, i.e., all vertices are removed.

This deterministic algorithm can be readily encoded in \dlv. The idea is to
use stages for modeling the iterations through Steps~2--5. Since
the number of steps is bounded by the number of rules in $\p$, we reuse
rule-IDs as stages:

\smallskip

\begin{tprogram}
\clause{stage(T)}{rule(T)}
\end{tprogram}

\smallskip

Stages are ordered by \dlv's built-in order $<$ on
constants. The first (least) stage is used for the stage after
the first run through Steps~2--5.

We use predicates $g$ and $z$ for rule labels ``g'' and ``z'',
 respectively, which are defined as follows (label ``i'' is not of
 interest and thus omitted):

\smallskip

\begin{tprogram}
\clauseB{g(R)}{rule(R), pos\_body\_true(R),}\\
\clauseE{\naf neg\_body\_false(R)}\\

\clauseB{z(R)}{rule(R), pos\_body\_true(R),}\\
\clauseE{head(X,R), \naf in\_AS(X)}
\end{tprogram}

\smallskip

Initially, only sources which are not zombies can be removed from the
graph. We use a predicate $nosource0(R)$, which informally means that
$R$ is {\em not} a source node in $G$, and a predicate $remove(R,S)$ which means that
at stage $S$, the vertex $R$ is no longer in $G$:

\smallskip

\begin{tprogram}
\clause{nosource0(R)}{pr(R1,R)}\\
\clauseB{remove(R,S)}{rule(R), \naf nosource0(R),}\\
\clauseE{\naf z(R), stage(S)}
\end{tprogram}

\smallskip

At other stages of the iteration, we can remove all rules satisfying
the condition of Step~3. We use a predicate $nosource(R,S)$ which
expresses that $R$ is not a source at stage $S$. 

\smallskip

\begin{tprogram}
\clauseB{nosource(R,S)}{pr(R1,R), stage(S),}\\
\clauseE{\naf remove(R1,S)}\\
\clauseB{remove(R,S1)}{rule(R), \naf nosource(R,S),}\\
\clauseM{stage(S), stage(S1), S < S1,}\\
\clauseE{\naf z(R), \naf remove(R,S)}\\
\clauseB{remove(R,S1)}{rule(R), \naf nosource(R,S),}\\
\clauseM{time(S), time(S1), S < S1,}\\
\clauseE{z(R), nbl(X,R), s(X,S)}
\end{tprogram}

\smallskip

According to Step~4, we must add the head $H(r)$ of a generating rule
which is to be removed in Step~5, to the set $S$ there. We represent
this using a predicate $s(X,St)$, which informally means that $X$
belongs to set $S$ at stage $St$, and add the rule:

\smallskip

\begin{tprogram}
\clause{s(X,St)}{remove(R,St), g(R), head(X,R)}
\end{tprogram}

\smallskip

Finally, according to Step~2 we have to check whether all rules
have been removed in the processing of the graph $G$. This is done
by using a predicate $removed$ for the projection of $remove$ to rules
and the following rule plus a constraint:

\smallskip

\begin{tprogram}
\clause{removed(R)}{remove(R,S)}\\
\clause{}{rule(R), \naf removed(R)}
\end{tprogram}

\smallskip

The resulting meta-interpreter program \PIg (without \PIa) is shown in Figure~\ref{fig:PIg}. Note
that \PIg is in general also more efficient than \PIb, since
unnecessary totalizations of the partial order can be avoided with
\PIg.
By virtue of the results in \cite{brew-eite-99} (in particular,
Lemma~7.2 there), we can state the follwing result:

\begin{figure}
{\small
\begin{alltt}
% Label 'g' nodes and 'z' nodes (other labels are uninteresting): 
  g(R) :- rule(R), not neg_body_false(R), pos_body_true(R). 
  z(R) :- rule(R), pos_body_true(R), head(X,R), not in_AS(X).

% Use rules ids as stages. 
  stage(S) :- rule(S). 

% Initial step of the algorithm: Consider global source nodes.
% Only non-z nodes can be removed.
  nosource0(R) :- pr(R1,R). 
  remove(R,S)  :- rule(R), not nosource0(R), not z(R), stage(S). 

% Other steps in the algorithm: Remove non-z nodes and, under some
% conditions, also z-nodes.
  nosource(R,S) :- pr(R1,R), stage(S), not remove(R1,S). 
  remove(R,S1)  :- rule(R), not nosource(R,S), stage(S), stage(S1),
                   S < S1, not z(R), not remove(R,S).
  remove(R,S1)  :- rule(R), not nosource(R,S), stage(S), stage(S1),
                   S < S1, z(R), nbl(X,R), s(X,S).

% Add the head of a removed generating rule to the set S.
  s(X,St) :- remove(R,St), g(R), head(X,R). 

% Check whether all rules are removed.
  removed(R) :- remove(R,S). 
  :- rule(R), not removed(R). 
\end{alltt}
}
\caption{Meta-Interpreter \PIg for B-Preferred Answer Sets Using Deterministic Preferredness Checking (without \PIa)} 
\label{fig:PIg}
\end{figure}

\begin{theorem}
\label{theo:3}
Let $\p = (\P,<)$ be a propositional prioritized program. Then, (i) if $A \in \AS(\PIg
\cup F(\p))$ then $\pi(A) \in \BPAS(\p)$, and (ii) for each $A \in
\BPAS(\p)$, there exists some $A' \in \AS(\PIg \cup F(\p))$ such
that $\pi(A') = A$.
\end{theorem}

\begin{proof}[Proof (sketch)]
\noindent (i)\ As in the proof for Theorem~\ref{theo:2} we employ the notion of splitting a program  \cite{lifs-turn-94}. Let $Q=\PIg\cup F(\p)$ and
 $Q_a = \PIa\cup F(\p)$. Then for each answer set $A$ of $Q$,
 its restriction $A_a$ to the predicates of $Q_a$ is an answer set of
 $Q_a$ and $A \in \AS((\PIb\setminus Q_a) \cup A_a)$.
 By Theorem~\ref{theo:1}, $T=\pi(A_a)=\pi(A)$ is an
 answer set of $P$.
 
 We can now loosely argue that the deterministic variant of FULL-ORDER
 with input $T$ creates at most $n$ (where $n = |P|$) intermediate values for the set
 $S$ there (not counting the initialisation to $\emptyset$) referred to as
 $S_1,\ldots,S_n$, and implicitly creates (cumulative) sets
 $R_1,\ldots,R_n$ of removed rules.  It can be seen that there is a
 one-to-one mapping of rule labels $r_1,\ldots,r_n$, ordered by the
 \dlv built-in $<$, to $S_1,\ldots,S_n$ and $R_1,\ldots,R_n$ via
 \code{stage}. Now, $R_1 = \{ r \mid remove(r,r_1) \in A\}$ and $S_1 =
 \{ h \mid s(h,r_1) \in A\}$ by the definition of \code{nosource0},
 \code{remove} and \code{s}. We can proceed by induction and assume
 that $R_i = \{ r \mid remove(r,r_i) \in A\}$ and $S_i = \{ h \mid
 s(h,r_i) \in A\}$ for $1 \leq i < n$. Then, it can be seen that
 $R_{i+1} = \{ r \mid remove(r,r_{i+1}) \in A\}$ and $S_{i+1} = \{ h
 \mid s(h,r_{i+1}) \in A\}$ hold by definition of the predicates
 \code{nosource0}, \code{remove} and \code{s}. Observe further that
 $R_n = \{ r \mid removed(r) \in A\}$, and that the graph $G$ is empty
 iff $R_n = P$. Since $A$ satisfies the final constraint in \PIg, $R_n
 = P$ is guaranteed to hold. Therefore the algorithm outputs ``yes''
 and so $\pi(A) \in \BPAS(\p)$ holds.

\noindent (ii)\ For $A \in \BPAS(\p)$ we can construct an answer set
$A'$ of $Q$, such that (again by notion of splitting) $A''$ is the
restriction of $A'$ to the predicates defined by $Q_a=\PIa \cup F(\p)$ and $A'$ is an
answer set of the program $Q_1$, which
contains $A''$ plus all clauses $c$ in $\Ground{Q}$ which involve the
predicates \code{g}, \code{z}, \code{nosource0}, \code{remove},
\code{s}, \code{removed} such that $A''$ satisfies all literals in $B(c)$ involving
predicates defined in $Q_a$. $Q_1$ is locally
stratified by a stratification defined as follows (where $r_1,\ldots,r_n$ are the rule identifiers ordered by \dlv's built-in $<$):
\begin{itemize}
\item $\lambda(\code{removed(r)}) = 2 \times n + 1$
\item $\lambda(\code{nosource(r,r_i)}) = 2 \times i + 1$
\item $\lambda(\code{remove(r,r_i)}) = \lambda(\code{s(h,r_i)}) = 2 \times i$
\item $\lambda(g(r)) = \lambda(z(r)) = 1$
\item $\lambda(a) = 0$ for all other atoms $a$ in $Q_1$
\end{itemize}
Since $\pi(A'')=\pi(A')=A$ must hold, it is easy to see that $A'$ is
an extension to an $A''$ (which must be an answer set of $\PIa \cup
F(\p)$) that is fully determined by $\lambda$. $A'$ must furthermore
satisfy the final constraint in $\PIg$. Due to these facts, such an
$A'$ can be effectively constructed.
\end{proof}

\begin{example}
  Consider the program in Example~\ref{exa:weak} and assume priorities
  $(1)<(3)$, $(2)<(4)$, and $(4)<(3)$. Suppose preferredness of
  $A_2=\{c,\neg d\}$ is checked. Then, the atoms $z(r1)$, $g(r2)$, and
  $g(r3)$ representing labels are derived, as well as $nosource0(r4)$
  and $nosource0(r3)$. Both $r1$ and $r2$ are sources, but $r1$ is
  labeled ``$z$'', so only $remove(r2,ri)$ and $s(c,ri)$ is derived
  for $i=1,\ldots,4$. Thus, $nosource(ri,r1)$ is derived only for
  $i=3$. Since $s(c,r1)$ holds, too, we can derive $remove(r1,ri)$ and
  $remove(r4,ri)$ for $i=2,3,4$. Neither $s(a,ri)$ nor $s(b,ri)$ are
  derived since $g(r1)$ and $g(r4)$ do not hold. Finally,
  $remove(r3,ri)$ and $s(\neg d,ri)$ for $i=3,4$ are derived and
  $removed(ri)$ holds for $i=1,\ldots,4$, satisfying the final
  constraint introduced above. Thus, $A_2$ is a preferred answer set.
\end{example}

An alternate definition of B-preferred answer sets is provided by
\cite{scha-wang-2001}; a meta-interpreter program following that definition
can be developed using techniques similar to the ones employed in
$\PIb$, $\PIg$, and the interpreters in the following sections.

\subsection{W-Preferred Answer Sets}
\label{sec:w-pref}

The semantics we have seen in Section~\ref{sec:pas} is but one way to
assign a meaning to prioritized logic programs. In this section we will
introduce a related approach due to Wang, Zhou and Lin
\cite{wang-etal-2000} following the presentation in \cite{scha-wang-2001}.

\begin{definition}[W-preferred answer set, $\WPAS$]
\label{def:w-pref}
Let $\p = \{ r_1,\ldots,r_n\}$ be a prioritized program. 
For any set $S \subseteq \BP$ of literals, the sequence $S_i\subseteq \BP$
($0 \leq i \leq n$) is defined as follows:
$$
\begin{array}{ll}
S_0= & \emptyset \\[1.2ex]
S_i= & S_{i-1} \cup \left\{ H(r) \left| \begin{array}{rp{0.50\textwidth}}
    \textrm{I.} & $r \in \p$ is active wrt.\ $(S_{i-1},S)$, and\\
   \textrm{II.} & there is no rule $r' \in \p$ with $r' < r$ such that \\
                & (a) $r'$ is active wrt.\ $(S,S_{i-1})$, and \\
                & (b) $H(r') \not\in S_{i-1}$.
  \end{array}
  \right.\right
  \}
\end{array}
$$
for all $i=1,\ldots,n$, where a rule $r$ is \emph{active} wrt.\ the pair
$(X,Y)$ if $\BpR \subseteq X$ and $\BnR \cap Y = \emptyset$.

The set $\Cw{\p}(S)$ is defined by 
$$
\Cw{\p}(S) =\left\{
\begin{array}{lp{0.30\textwidth}}
 S_n  & if $S_n$ is consistent, \\
 \BP{} & otherwise.
\end{array}\right.
$$

and $S$ of $\p$ is W-preferred if $\Cw{\p}(S) = S$. The set of
all W-preferred answer sets of $\p$ is denoted by $\WPAS(\p)$. In this paper
we will only consider consistent W-preferred answer sets.
\end{definition}  

\subsubsection{Adapting the Meta-Interpreter}

\begin{figure}
{\scriptsize
\begin{alltt}
% Part 1: Guess a consistent set S.

  lit(L) :- head(L,_).   lit(L) :- bpl(L,_).   lit(L) :- nbl(L,_).

  in_S(L) v notin_S(L) :- lit(L).
  :- compl(X,Y), in_S(X), in_S(Y).

% Part 2: Handle preferences.

  pr(X,Z) :- pr(X,Y), pr(Y,Z).
  :- pr(X,X).

% Part 3: Stage IDs.

  stage(S) :- rule(S).

% Part 4: Evaluate positive bodies.

  pos_body_false_S(Y) :- rule(Y), pbl(X,Y), not in_S(X).
  pos_body_false_Si(R,Si) :- pbl(L,R), stage(Si), not in_Si(L,Si).
  pos_body_false_S0(R) :- pbl(L,R).

% Part 5: Evaluate negative bodies.

  neg_body_false_S(Y) :- rule(Y), nbl(X,Y), in_S(X).
  neg_body_false_Si(R,Si) :- nbl(L,R), stage(Si), in_Si(L,Si).

% Part 6: Determine active rules.

  active(R,Si) :- rule(R), stage(Si),
                  not pos_body_false_Si(R,Si), not neg_body_false_S(R).

  active_Si(R,Si) :- rule(R), stage(Si), not pos_body_false_S(R),
                     not neg_body_false_Si(R,Si).
  active_S0(R) :- rule(R), not pos_body_false_S0(R), not neg_body_false_S(R).

% Part 7: Check for preferred generating rules.

  head_not_in_Si(R,Si) :- stage(Si), head(H,R), not in_Si(H,Si).

  preferred_generating_rule_exists(R,Si) :- pr(R1,R), active_Si(R1,Si),
                                            head_not_in_Si(R1,Si).
  preferred_generating_rule_exists_S0(R) :- pr(R1,R), not pos_body_false_S(R1).

% Part 8: Compute Si.

  in_Si(H,Si) :- head(H,R), active(R,Sj), stage(Sj), stage(Si), Si > Sj,
                 not preferred_generating_rule_exists(R,Sj).
  in_Si(H,Si) :- head(H,R), active_S0(R), stage(Si),
                 not preferred_generating_rule_exists_S0(R).

% Part 9: Verify "stability".

  in_PAS(L) :- in_Si(L,_).
  :- in_PAS(L), not in_S(L).
  :- in_S(L), not in_PAS(L).
\end{alltt}
}
\caption{Meta-Interpreter \PIw for W-Preferred Answer Sets} 
\label{fig:PIw}
\end{figure}

In Figure~\ref{fig:PIw} we provide a meta-interpreter for W-preferred
answer sets which closely follows Definition~\ref{def:w-pref} and
consists of three parts: The first guesses a consistent literal set
$A$ (Part 1 below), the second proceeds in stages of rule application
according to the definition (Parts 2--8 below), and the final one
verifies the ``stability'' condition $\Cw{\p}(A) = A$ (Part 9).

\paragraph{Part 1 [Guess a consistent set S]}
By means of the first three rules we extract all literals occurring in
the input program $\p$ into a new predicate \code{lit}. Then we guess
all possible subsets $S$ of \code{lit} by means of the disjunctive
rule such that \code{in\_S(X)} is true iff $\code{X} \in S$. The constraint,
finally, ensures that the set $S$ is consistent.

\paragraph{Part 2 [Handle Preferences]}
To complete the preference relation we transitively close the
\code{pr} predicate and we also verify that it is irreflexive.
The constraint is violated (in that case $\PIw \cup F(\p)$ admits
no answer set) only if a rule is preferred to itself -- \code{pr(X,X)}.

\paragraph{Part 3 [Stage IDs]}
Similar to Definition~\ref{def:w-pref} where we have used the indices
of the rules $r_1,\ldots,r_n$, we reuse the IDs of the rules in $\p$
and the built-in arbitrary order $<$ over these IDs as IDs of the
consecutive stages $S_1, S_2,\ldots$ of the definition. We can safely
do that, as the number of rules is an upper bound for the number of
stages of the computation of $\PIw$.

\paragraph{Part 4 [Evaluate positive bodies]}
According to Definition~\ref{def:w-pref} we need to evaluate the positive
and negative bodies of the rules in $\p$ in two ways to verify whether a
rule is active wrt.\ $(S_i, S)$ and $(S, S_i)$, respectively.

The predicates \code{pos\_body\_false\_S(R)} and
\code{pos\_body\_false\_Si(R,Si)} represent the sets of all rules \code{R}
whose bodies are false according to the set $S$ we have guessed in
Part~1 and the set $S_i$, respectively, where $S_i$ is represented by
the predicate \code{in\_Si} (with $S_i = \{ \code{L} \ |\
\code{in\_Si(L,Si)} \}$).

\code{pos\_body\_false\_S0(R)} covers the base case for $S_0 = \emptyset$
where the positive body is false w.r.t. $S_0$ if some positive body
literal exists.

\paragraph{Part 5 [Evaluate negative bodies]}
This works analogously to the case of positives bodies, just that we do
not need (and thus omit) the special case for $S_0$, as no negative body literal
can occur in $S_0 = \emptyset$.

\paragraph{Part 6 [Determine active rules]}
Now we need to define those rules that are active wrt.\ $(S_i,S)$ 
and $(S,Si)$. The former is handled by the first rule, the latter
by the second and third rules, where $S$ and $S_i$ are represented
by \code{in\_S} defined in Part~1 and $S_i$ is represented by the
predicate \code{in\_Si} that we will define in the following.
Again, the third rule covers the special case for the initial stage
$S_0 = \emptyset$.

\paragraph{Part 7 [Check for preferred generating rules]}
The rule with %
\code{preferred\_generating\_rule\_exists(R,Si)} in its head checks whether a
rule, which is preferred to R, exists such that it is active wrt.\
$(S,S_i)$ and its head does not occur in $S_i$ (where
\code{Si} represents $S_i$).

\code{head\_not\_in\_Si(R,Si)} here is used as an auxiliary predicate
that checks whether the head of the rule \code{R} is in the set $S_i$.

The third rule once more covers the base case $(S, S_0) = (S, \emptyset)$,
where we can simplify the body of the rule as shown.

\paragraph{Part 8 [Compute $S_i$]}
To compute $S_i$, we have to include the head of all rules which are
active wrt.\ $(S_{i-1},S)$ and where no preferred rule exists which is
active wrt.\ $(S,S_{i-1})$ and whose head does not already occur in
$S_{i-1}$.
Also here we need a specialized rule for the base-case where $S_{i-1} =
S_0$.

\paragraph{Part 9 [Verify ``stability'']}
Finally, we define a predicate \code{in\_PAS} as the union of $S_i$,
for $1 \leq i \leq n$, and check the ``stability'' condition of 
Definition~\ref{def:w-pref}, i.e., we check whether the relations
\code{in\_S} and \code{in\_PAS} are equal. Any difference between these
two will lead to a violation of one of the two constraints and thus
a corresponding answer set for $\PIw \cup F(\p)$ cannot exist.

\bigskip

We provide the following theorem that states the correctness of the
meta-interpreter program \PIw, where $\pi'(S) = \{ \ell \mid
in\_PAS(\ell) \in S\}$.

\begin{theorem}
\label{theo:4}
  Let $\p = (\P,<)$ be a propositional prioritized program. Then, (i)
  if $A \in \AS(\PIw \cup F(\p))$ then $\pi'(A) \in \WPAS(\p)$, and (ii)
  for each $A \in \WPAS(\p)$, there exists some $A' \in \AS(\PIw \cup
  F(\p))$ such that $\pi'(A') = A$.
\end{theorem}

\begin{proof}[Proof (sketch)]
  For the proof of the result the same techniques as in the proofs of Theorems~\ref{theo:2} and
  \ref{theo:3} can be used. The major difference is that \PIw does not start
  from answer sets generated by \PIa, but from consistent sets
  generated by the rules in Part~1 of \PIw. Therefore the splitting is
  done on the literals defined by the rules in Part~1 and $F(\p)$.
  
  As for (i), it can be shown inductively that the sets $S_j$ ($1 \leq
  j \leq n$) of Definition~\ref{def:w-pref} correspond to the sets $\{ h \mid in\_Si(h,r_j)\}$ where $n = |P|$ is again an upper bound and
  the $r_i$ are again rule labels ordered by the \dlv built-in order.
  The constraints in Part~9 guarantee that the criterion $\Cw{\p}(S) =
  S$ is met.
  
  As for (ii), the split program $Q_1$ (obtained from the ground program by dropping rules from Part~1 and
  $F(\p)$ and adding an answer set $A''$ of the dropped rules, while keeping only those rules of which the bodies agree with $A''$) is again
  locally stratified. A possible stratification would be as follows:
\begin{itemize}
\item $\lambda(\code{in\_PAS(l)}) = 3 \times n+2$
\item $\lambda(\code{preferred\_generating\_rule\_exists(r,r_i)}) = 3 \times i + 2$
\item $\lambda(\code{active(r,r_i)}) = 3 \times i + 2$
\item $\lambda(\code{head\_not\_in\_Si(r,r_i)}) = 3 \times i + 1$
\item $\lambda(\code{active\_Si(r,r_i)}) = 3 \times i + 1$
\item $\lambda(\code{pos\_body\_false\_Si(r,r_i)}) = 3 \times i + 1$
\item $\lambda(\code{neg\_body\_false\_Si(r,r_i)}) = 3 \times i$
\item $\lambda(\code{in\_Si(h,r_i)}) = 3 \times i$
\item $\lambda(\code{preferred\_generating\_rule\_exists\_S0(r)}) = 2$
\item $\lambda(\code{pos\_body\_false\_S(r)}) = 1$
\item $\lambda(\code{active\_S0(r)}) = 1$
\item $\lambda(a) = 0$ for all other atoms $a$ in $Q_1$
\end{itemize}
  Again, based on this information we can effectively construct an $A'
  \in \AS(\PIw \cup F(\p))$ such that $\pi'(A') = A$.
\end{proof}

\subsection{D-Preferred Answer Sets}
\label{sec:d-pref}

Another way to assign a meaning to prioritized logic programs has been
introduced by Delgrande, Schaub and Tompits \cite{delg-etal-00a}. For
our presentation we again follow \cite{scha-wang-2001}.

\begin{definition}[D-preferred answer set, $\DPAS$]
Let $\p = \{ r_1,\ldots,r_n\}$ be a prioritized program. 
For any set $S \subseteq \BP$ of literals, the sequence $S_i\subseteq \BP$
($0 \leq i \leq n$) is defined as follows:
$$
\begin{array}{ll}
S_0= & \emptyset \\[1.2ex]
S_i= & S_{i-1} \cup \left\{ H(r) \left| \begin{array}{rp{0.50\textwidth}}
    \textrm{I.} & $r \in \p$ is active wrt.\ $(S_{i-1},S)$, and\\
   \textrm{II.} & there is no rule $r' \in \p$ with $r' < r$ such that \\
                & (a) $r'$ is active wrt.\ $(S,S_{i-1})$, and \\
                & (b) $r' \not\in rule(S_{i-1})$.
  \end{array}
  \right.\right
  \}
\end{array}
$$
for all $i=1,\ldots,n$, where again a rule $r$ is \emph{active} wrt.\ the
pair $(X,Y)$ if $\BpR \subseteq X$ and $\BnR \cap Y = \emptyset$, and
$rule(X)$ denotes those rules $\in \p$ that have been effectively used
to derive literals $\in X$.

The set $\Cd{\p}(S)$ is defined by 
$$
\Cd{\p}(S) =\left\{
\begin{array}{lp{0.30\textwidth}}
 S_n,  & if $S_n$ is consistent, \\
 \BP{} & otherwise.
\end{array}\right.
$$

and an answer set $A$ of $\p$ is D-preferred if $\Cd{\p}(A) = A$. The set of
all D-preferred answer sets of $\p$ is denoted by $\DPAS(\p).$
\end{definition}  

The basic difference between D-preferred and W-preferred answer sets
is that the former requires that a higher-ranked rule has been used to
actually derive some literal, while for the latter it is sufficient that
the literal appears in the head of such a rule.

In fact one can show that the three approaches we have shown get
increasingly restrictive in that each approach admits only a subset
of the (preferred) answer sets of the previous approach. The following
theorem is due to \cite{scha-wang-2001}:

\begin{theorem}[Schaub \& Wang, 2001] %
Let $\p = (\P,<)$ be a 
propositional prioritized program.  Then, we have:
$$
\DPAS(\p) \subseteq \WPAS(\p) \subseteq \BPAS(\p) \subseteq \AS(\p)
$$
\end{theorem}

\subsubsection{Adapting the Meta-Interpreter}

\begin{figure}
{\scriptsize
\begin{alltt}
% Part 1: Guess a consistent set S.

  lit(L) :- head(L,_).   lit(L) :- bpl(L,_).   lit(L) :- nbl(L,_).

  in_S(L) v notin_S(L) :- lit(L).
  :- compl(X,Y), in_S(X), in_S(Y).

% Part 2: Handle preferences.

  pr(X,Z) :- pr(X,Y), pr(Y,Z).
  :- pr(X,X).

% Part 3: Stage IDs.

  stage(S) :- rule(S).

% Part 4: Evaluate positive bodies.

  pos_body_false_S(Y) :- rule(Y), pbl(X,Y), not in_S(X).
  pos_body_false_Si(R,Si) :- pbl(L,R), stage(Si), not in_Si(L,Si).
  pos_body_false_S0(R) :- pbl(L,R).

% Part 5: Evaluate negative bodies.

  neg_body_false_S(Y) :- rule(Y), nbl(X,Y), in_S(X).
  neg_body_false_Si(R,Si) :- nbl(L,R), stage(Si), in_Si(L,Si).

% Part 6: Determine active rules.

  active(R,Si) :- rule(R), stage(Si),
                  not pos_body_false_Si(R,Si), not neg_body_false_S(R).

  active_Si(R,Si) :- rule(R), stage(Si), not pos_body_false_S(R),
                     not neg_body_false_Si(R,Si).
  active_S0(R) :- rule(R), not pos_body_false_S0(R), not neg_body_false_S(R).

% Part 7: Check for preferred generating rules.

| rule_not_generating_in_Si(R,Si) :- stage(Si), head(H,R),
|                                    not in_rule_Si(H,R,Si).

| preferred_generating_rule_exists(R,Si) :- pr(R1,R), active_Si(R1,Si),
|                                           rule_not_generating_in_Si(R1,Si).
  preferred_generating_rule_exists_S0(R) :- pr(R1,R), not pos_body_false_S(R1).

% Part 8: Compute Si.

| in_rule_Si(H,R,Si) :- head(H,R), active(R,Sj), stage(Sj), stage(Si), Si > Sj,
|                not preferred_generating_rule_exists(R,Sj).
| in_rule_Si(H,R,Si) :- head(H,R), active_S0(R), stage(Si),
|                not preferred_generating_rule_exists_S0(R).

| in_Si(H,Si) :- in_rule_Si(H,_,Si).

% Part 9: Verify "stability".

  in_PAS(L) :- in_Si(L,_).
  :- in_PAS(L), not in_S(L).
  :- in_S(L), not in_PAS(L).
\end{alltt}
}
\caption{Meta-Interpreter \PId for D-Preferred Answer Sets} 
\label{fig:PId}
\end{figure}

The changes from \PIw to \PId are relatively small, and we have marked
those lines where \PId differs by a vertical bar in Figure~\ref{fig:PId}.

Instead of tracking literals by means of \code{in\_Si} we need to track
which concrete rule has been used to derive a particular literal, and we
do this by means of a new predicate \code{in\_rule\_Si(H,R,Si)} which
specifies that in the state denoted by \code{Si} the literal \code{H}
has been derived by means of the rule \code{R}.

Similarly, we replace \code{head\_not\_in\_Si} by a new predicate
\code{rule\_not\_generating\_in\_Si} that considers whether a specific rule
has been actually applied, not just whether the head of this rule has been
derived (possibly from a different rule).

\code{in\_Si}, finally, is a simple projection of \code{in\_rule\_Si} to
obtain the union of all $S_i$s for use in the stability check.

We have the following result (recall that $\pi'(S) = \{ \ell \mid
in\_PAS(\ell) \in S\}$).

\begin{theorem}
\label{theo:6}
  Let $\p = (\P,<)$ be a propositional prioritized program. Then, (i)
  if $A \in \AS(\PId \cup F(\p))$ then $\pi'(A) \in \DPAS(\p)$, and (ii)
  for each $A \in \DPAS(\p)$, there exists some $A' \in \AS(\PId \cup
  F(\p))$ such that $\pi'(A') = A$.
\end{theorem}

\begin{proof}[Proof (sketch)]
  The proof sketched for Theorem~\ref{theo:4} can be adapted in a
  straightforward way.
\end{proof}

\section{Weakly Preferred Answer Sets}
\label{sec:wpas}

The concept of weakly preferred answer set relaxes the priority
ordering as little as necessary to obtain a preferred answer set, if
no answer set is preferred. It can be seen as a conservative
approximation of a preferred answer set. So far, this approximation
has only been defined for B-preferred answer sets \cite{brew-eite-99},
though similar extensions to the two other approaches we have seen
in Sections~\ref{sec:w-pref} and \ref{sec:d-pref} should be feasible.

\begin{definition}[distance]
\label{def:distance}
Let $<_1$ and $<_2$ be total orderings of the same finite set $M$. The
distance from $<_1$ to $<_2$, denoted $d(<_1,<_2)$, is the number of
pairs $m,m' \in M$ such that $m <_1 m'$ and $m' <_2 m$.%
\footnote{The definition in \cite{brew-eite-99} uses ordinals and deals with possibly infinite $M$. 
Ours is equivalent on finite $M$.}
\end{definition}

Clearly, $d(<_2,<_1)$ defines a metric on the set of all total
orderings of $M$. For example, the distance between $a <_1 b <_1 c$
and $c<_2 a <_2 b$ is $d(<_1,<_2)=d(<_2,<_1)=2$. Note that
$d(<_1,<_2)$ amounts to the smallest number of successive switches of
neighbored elements which are needed to transform $<_1$ into
$<_2$. This is precisely the number of switches executed by the
well-known bubble-sort algorithm.

\begin{definition}[preference violation degree, pvd]
\label{def:pvd}
Let $\p = (P,<)$ be a prioritized program and let $A \in \AS(\p)$.
The preference violation degree
of $A$ in $\p$, denoted $pvd_\p(A)$, is the minimum distance from any
full prioritization of $\p$ to any fully prioritized program $\p' =(P,<')$ such that $A$ is a preferred answer set of
$\p'$, i.e.,
$$
pvd_{\p}(A) = \min\{ d(<_1,<_2) 
                     \mid (P,<_1) \in \FP(\p), A \in \BPAS(P,<_2) \}.
$$
The preference violation degree of $\p$, $pvd(\p)$, is defined by
$pvd(\p) = \min \{ pvd_\p(A) \mid A \in \AS(P) \}$.
\end{definition}

Now the weakly preferred answer sets are those answer sets which minimize
preference violation.

\begin{definition}[weakly preferred answer set, $\weakPAS$]
\label{def:weakly_preferred_answer_set}
Let $\p = (P,<)$ be a prioritized program. Then, $A\in \AS(\P)$ is a
weakly preferred answer set of $\p$ iff $pvd_\p(A) = pvd(\p)$.
By $\weakPAS(\p)$ we denote the collection of all such weakly preferred
answer sets of $\p$.
\end{definition}

\begin{example} In the bird \& penguin example, $A_1$
is the unique preferred answer set of $\p$. Clearly, every preferred
answer set $A$ of any prioritized program $\p$ has $pvd_\p(A)=0$, and
thus $A$ is a weakly preferred answer set of $\p$.  Thus, $A_1$ is
the single weakly preferred answer set of the program.
\end{example}

\begin{example}
\label{exa:nonex-prans2}
Reconsider the program in Example~\ref{exa:nonex-prans}. Its answer
set $A= \{ b\}$ is not preferred. Switching the priorities of the two rules,
the resulting prioritized program $\p'$ has $\Cb{\dr{A}{\p'}}(A) =
\{b\}$, thus $A$ is preferred for $\p'$. Hence $pvd_{\p}(A) =
        pvd(\p)=1$ and
$A$ is a weakly preferred answer set of \p.
\end{example}

\begin{example}
\label{exa:weak}
Consider the following program $\p$:
\ms

\begin{tprogram}
(1) \quad\hfill \clause{a}{\naf c} \\ 
(2) \hfill \clause{c}{\naf b} \\ 
(3) \hfill \clause{\neg d}{\naf b} \\
(4) \hfill \clause{b}{\naf \neg b, a} \\
\end{tprogram}
\ms

$\p$ has the answer sets $A_1 = \{a,b\}$ and $A_2 = \{
c, \neg d\}$. Imposing $(i)<(j)$ iff $i<j$, none is preferred. We have
$pvd_{\p}(A_1) = 2$:\ (2) and (3) are zombies in the
dual reduct which are only defeatable by (4), which must be moved in front of them;
this takes two switches. On the other hand, $pvd_{\p}(A_2) =
1$: the single zombie (1) in the dual reduct is defeated if (2) is moved in
front of it (here, (4) is a dead rule). Hence,
$pvd(\p)=1$, and $A_2$ is the single weakly preferred answer set of
$\p$.
\end{example}

\subsection{Adapting the Meta-Interpreter for Weakly Preferred Answer Sets}

The transition from an interpreter for preferred answer sets to one
for weakly preferred answer sets is simple -- just a few
clauses have to be added and one has to be slightly altered.

For weakly preferred answer sets, we have to generate a second total ordering
(called $pr1$), which needs not be compatible with the input partial order, and corresponds to $<_2$ in definition~\ref{def:pvd}.
$$
\begin{tprogram}
\clause{pr1(X,Y) \Or pr1(Y,X)}{rule(X),\ rule(Y),\ X\, !\!\!= Y}\\
\clause{pr1(X,Z)}{pr1(X,Y),\ pr1(Y,Z)}\\
\clause{}{pr1(X,X)}
\end{tprogram}
$$
This ordering should be used to determine the preferred answer
sets. Since the given totalization of the input ordering occurs just
in one rule of the original program, we just have to update this rule:
$$
\begin{tprogram}
\clause{defeat\_local(Y)}{nbl(X,Y),\ lit(X,Y1),\ pr1(Y1,Y)}
\end{tprogram}
$$
Finally, we want to keep only those orderings which minimize the
differences to some totalization of an input ordering. To this end, we
state a weak constraint, where each difference in the orderings gets a
penalty of one (we don't need the leveling concept here).
\begin{tprogram}
\wclause{pr(X,Y), pr1(Y,X)}{1:1}
\end{tprogram}
In this way, each answer set $A$ will be weighted with $pvd_\p(A)$,
and the optimal answer sets minimize this number, which corresponds
exactly to Defs.~\ref{def:distance}, \ref{def:pvd}, and
\ref{def:weakly_preferred_answer_set}.

\begin{figure}[tb]
{\small
\begin{alltt}
% For full prioritization: Refine pr to a total ordering.
  pr(X,Y) v pr(Y,X) :- rule(X),rule(Y), X != Y. 
  pr(X,Z) :- pr(X,Y), pr(Y,Z).
  :- pr(X,X). 

% Weakly preferred answer sets: Create a total ordering pr1,
% as close to pr as possible.
  pr1(X,Y) v pr1(Y,X) :- rule(X),rule(Y), X != Y. 
  pr1(X,Z) :- pr1(X,Y), pr1(Y,Z). 
  :- pr1(X,X). 

% Weak constraint: Minimize violations.
  :~ rule(X), rule(Y), pr(X,Y), pr1(Y,X). [1:1]

% Check dual reduct: Build sets S_i, use rule ids as indices i.
% lit(X,r) means that the literal x occurs in the set S_r.
   lit(X,Y) :- head(X,Y), pos_body_true(Y), 
               not defeat_local(Y), not in_AS(X). 
   lit(X,Y) :- head(X,Y), pos_body_true(Y), 
               not defeat_local(Y), not defeat_global(Y). 
   defeat_local(Y)  :- nbl(X,Y), lit(X,Y1), pr1(Y1,Y).  
   defeat_global(Y) :- nbl(X,Y), in_AS(X). 

% Include literal into CP(.).
   in_CP(X) :- lit(X,Y).
   :- in_CP(X), not in_AS(X).
\end{alltt}
}
\caption{Meta-Interpreter \PIweak for Weakly Preferred Answer Sets (without \PIa)} 
\label{fig:PIweak}
\end{figure}

Let us call the resulting interpreter \PIweak; a compact listing (without showing \PIa) is
given in Figure~\ref{fig:PIweak}. We have the following result:

\begin{theorem}
Let $\p = (\P,<)$ be a propositional prioritized program. Then, (i)
if 
$A \in \OAS(\PIweak \cup F(\p))$, then $\pi(A) \in \weakPAS(\p)$, and
(ii) for each $A \in \weakPAS(\p)$, there exists some $A' \in
\OAS(\PIweak \cup F(\p))$ such that $\pi(A') = A$.
\end{theorem}

\begin{proof}
Let $Q_0$ be the set of all clauses in $\PIweak$ except the two rules
defining \code{pr}, the constraint \code{\derives pr(X,X).}, and the
weak constraint for minimization of violations. After renaming
\code{pr1} to \code{pr}, $Q_0$ is identical to the meta-interpreter
program $\PIb$ minus the redundant constraint \code{\derives
in\_AS(X), \naf in\_CP(X)}. Thus, we infer from Theorem~\ref{theo:2}
that $\AS(Q_0\cup F(P,\emptyset))$ is in correspondence
(i), (ii) as there to $\BPAS(P,\emptyset)$. Let $Q_1$ result from $Q_0$ by adding the rules defining
\code{pr} and the constraint \code{\derives pr(X,X).} Then, $\AS(Q_1\cup F(\p))$
is in a similar correspondence to the set of tuples $T = \{
(A,<_1,<_2) \mid$ $(P,<_1)\in \FP(\p)$, $A \in \BPAS(P,<_2)$,
$(P,<_2)\in \FP(P,\emptyset)\}$. Adding the weak constraint to $Q_1$ (which results in $\PIweak$), we thus have
that $\OAS(\PIweak\cup F(\p))$ is in similar
correspondence to the set $T' = \{ (A,<_1,<_2)\in T \mid$ $\forall\,
(A',<'_1,<'_2) \in T: d(<'_1,<'_2)\geq d(<_1,<_2)\,\}$, which in turn
naturally corresponds to $\weakPAS(\p)$. 
More precisely, we can conclude that for each $A\in \OAS(\PIweak \cup
F(\p))$, there exists some tuple $(\pi(A),<_1,<_2)$ in $T'$,
which corresponds to $\pi(A) \in \weakPAS(\p)$; conversely, for each
$A\in \weakPAS(\p)$, there exists some tuple $(A,<_1,<_2) \in T'$,
which corresponds to some $A'\in \OAS(\PIweak \cup F(\p))$ such that
$\pi(A')=A$. This proves items (i) and (ii) of the theorem.
\end{proof}

\begin{example}
Reconsider Example~\ref{exa:nonex-prans}, which has
no preferred answer set. $\PIweak \cup F(\p)$ has one optimal answer set
  (with weight 1 in level 1) containing $in\_AS(b)$, $pr(r1,r2)$, and
  $pr1(r2,r1)$, which is consistent with Example~\ref{exa:nonex-prans2}.
\end{example}

\begin{example}
  Reconsider Example~\ref{exa:weak}, which does not have any preferred
  answer set either. $\PIweak \cup F(\p)$ has one optimal answer set
  (with weight 1 in level 1) containing $in\_AS(c)$, $in\_AS(\neg d)$,
  $pr(r1,r2)$, and $pr1(r2,r1)$, where the pair $(r1,r2)$ is the only
  difference between $pr$ and $pr1$, consistent with
  Example~\ref{exa:weak}.
\end{example}

While \PIweak is a straightforward encoding of the definition of
weakly preferred answer set, and gives us an executable specification,
it is quite inefficient on larger problem instances because of the
large search space generated by the possible total orderings $pr$ and
$pr1$.  To improve efficiency, we can use a variant of the graph algorithm
FULL-ORDER from Section~\ref{sec:det-checking},
based on the following observation. We
may arrange the vertices which are removed from $G$, in this order, as
a common prefix for orderings $<_1$ and $<_2$ in the definition of
$pvd(A)$.  We thus need to guess only optimal completions of $<_1$ and
$<_2$ for those rules that remain in $G$ on termination of FULL-ORDER.
In particular, if $G$ is empty, then $<_1$ and $<_2$ coincide and $A$
is a preferred answer set, hence also weakly preferred. The
meta-interpreter programs $\PIg$ and $\PIweak$ can be combined to another
meta-interpreter program for computing weakly  preferred answer sets, which
conservatively extends the computation of preferred answer sets in the
sense that guessing comes only into play if no preferred answer sets
exist. However, we do not further discuss this here.

\section{Related Work} 
\label{sec:rel-work}

Meta-interpretation of answer sets or answer set-like semantics has
been considered by other authors as well, in different contexts. We
briefly discuss
\cite{gelf-son-97,mare-remm-01,delg-etal-00a,delg-etal-01} which are
related to our work.

\paragraph{Gelfond and Son.} In \cite{gelf-son-97}, the idea of
meta-interpretation was used to define the semantics of a language
${\cal L}_0$ for rules with preferences.  ${\cal L}_0$ is a
multi-sorted logical language which has constants for individuals,
definite rules, and default rules of the
form ``If $l_1,\ldots,l_m$ are true, then normally $l_0$ is true,''
functions and relations for the domain as well as special predicates
for defining rules and expressing preference. For example, the formula
$$
default(d,l_0,[l_1,\ldots,l_m])
$$ 
represents a default rule, where $d$ is its name and $[l_1,\ldots,l_m]$
is Prolog-like list notation. Informally, it amounts to the rule $l_0
\derives l_1,\ldots, l_m,\naf\tneg l_0$ in extended logic programming.
Moreover, the language allows to express conflicts between two default
rules; both preferences and conflicts can be declared
dynamically by means of rules.

The semantics of ${\cal L}_0$ is then defined in terms of a
transformation of any program $\P$ in ${\cal L}_0$ into an extended
logic program $t(\P)$ whose answer sets are, roughly speaking, cast
into answer sets of the program $\P$.

However, there are some salient differences w.r.t.\ the approach of
\cite{gelf-son-97} and the one presented here.

\begin{itemize} 
\item First and foremost, the semantics of ${\cal L}_0$ is defined only by means of a
meta-interpreter, while our approach implements semantics which have
been defined previously without meta-interpretation techniques.

\item Secondly, the interpretation program in \cite{gelf-son-97} uses
lists for representing aggregations of literals and conditions on
them, in particular ``for all'' conditions. Such lists cannot be used
in datalog programs, as arbitrarily deep function nesting is required
for the list concept. We avoid these aggregations by using rule
identifiers, a traversal mechanism that exploits an implicit
ordering, and default negation.  

\item Thirdly, in our approach we extend a general answer set
meta-interpreter, thus clearly separating the representation of answer
sets and prioritization. In the meta-interpreter presented in
\cite{gelf-son-97}, this distinction is not obvious.
\end{itemize}

\paragraph{Marek and Remmel.} In a recent paper
\cite{mare-remm-01}, Marek and Remmel discussed
the issue of a meta-interpreter for propositional normal logic programs
in the context of the expressiveness of stable logic programming.
They describe a function-free normal logic program $\mathit{Meta1}$, such
that on input of a factual representation $edb_Q$ of a 0-2 normal
logic program $Q$ (i.e., each clause in $Q$ has 0 or 2 positive body
literals), a projection of the answer sets of $\mathit{Meta1} \cup
edb_Q$ is in one-to-one correspondence to the answer sets of $Q$.  The
representation $edb_Q$ is similar to our representation $F(Q)$, but
explicitly records the position of positive body literals.  The
meta-interpreter $\mathit{Meta1}$ is similar to ours, but differs
from ours in the following respects:

\begin{itemize}
\item Firstly, $\mathit{Meta1}$ encodes a simple guess and check
strategy for the computation of a stable model $S$. It contains a pair
of unstratified rules which guess for each atom $a$, whether $a$ is in
the stable model $S$ or not.  The remaining clauses mimic the
computation of the minimal model of $Q^S$, using a special predicate
$computed$, and constraints check whether $S$ is reconstructed by
it. On the other hand, our meta-interpreter \PIa has no separate
guessing and checking parts. Instead, stability of a model is
effected by the stable semantics underlying the
interpreter. Furthermore, \PIa uses negation sensible to the structure
of the program $Q$, and in essence preserves stratification (cf.\
Proposition~\ref{prop:strat}).

\item Secondly, because of its naive guess and check strategy,
$\mathit{Meta1}$ is highly inefficient for programs $Q$ which can be
evaluated easily. In particular, even for positive programs $Q$,
$\mathit{Meta1}$ explores an exponential search space, and computation
of the unique stable model of $Q$ may take considerable time. On the
other hand, for stratified $Q$
our meta-interpreter program \PIa is, after propagation of the input
facts, a locally stratified program and can be
evaluated efficiently. Loosely speaking, \PIa interprets a
significant class of computational ``easy'' logic programs
efficiently.

\item Thirdly, $\mathit{Meta1}$ is only applicable for 0-2 normal logic
programs. An extension to arbitrary normal logic programs is possible
using similar techniques as in this paper, though. 

\end{itemize}

All these considerations suggest the conclusion that the
meta-interpreter $\mathit{Meta1}$ in \cite{mare-remm-01} is more of
theoretical interest, which is fully compliant with the goals of that
paper.

\paragraph{Delgrande, Schaub, and Tompits.} Based on a seminal
approach for adding priorities to default logic \cite{delg-scha-97},
Delgrande {\em et al.} have developed the PLP framework for expressing
priorities on logic programs \cite{delg-etal-00a,delg-etal-01}. In
this framework, extended logic programs with preference information $r
\prec r'$ between rules attached at the object level, are ``compiled''
into another extended logic program, such that the answer sets of the
latter program amount to the preferred answer sets of the original
program.  The transformation uses a number of control predicates for
the application of rules such that rule preferences are respected as
intended in the application of the rules for constructing an answer
set of a given program. More specifically, control atoms $ap(r)$ and
$bl(r)$ state whether a rule $r$ is applied or blocked, respectively,
and atoms $ok(r)$ and $rdy(r)$ control the applicability of rules
based on antecedent conditions reflecting the given order
information. The framework provides the flexibility to modify the
standard transformation, such that transformations for different
preference semantics can be designed.

The PLP framework significantly differs from our work in the following
respects:

\begin{itemize}
\item Firstly, PLP does not use a fixed meta-interpreter for
evaluating varying programs, given in a format which can be
``processed'' by a meta-interpreter. Rather, PLP performs a schematic
program construction ad-hoc.

\item Secondly, PLP aims at a tool for realizing preferences
semantics by providing a suite of special predicates and a particular
representation formalism. In contrast, our interest is in casting
definitions from first principles to extended logic programs, in a way
such that we obtain executable specifications. This way, variations of
definitions can be experimented with more flexibly. 

\item Thirdly, similar as \cite{gelf-son-97}, PLP has no obvious
separation of answer sets and prioritization.
\end{itemize}

\section{Conclusion}
\label{sec:conclusion}

In this paper, we have considered the issue of building experimental
prototypes for semantics of extended logic programs equipped with
rule preferences, by using the technique of meta-interpretation. In
the course of this, we have presented a suite of meta-interpreters for
various such semantics, including a simple meta-interpreter for answer
set semantics of plain extended logic programs itself. This
meta-interpreter has benign computational properties, and can be used
as a building block for meta-interpreters of other semantics.

While the focus of this paper has been on a propositional setting, it
is possible to extend the techniques that we have presented for
handling non-ground programs as well. However, unless function symbols
are allowed at the code level (which is currently not the case in
\dlv), a technical realization is not completely
straightforward. Extending our work to deal with such cases, for which
the work reported in \cite{bona-01,bona-01a} might prove useful, and
creating a \dlv{} front-end for prioritized program evaluation are
issues for further work.

We believe that the work that we have presented in this paper provides
supportive evidence to the following items.

\begin{itemize}
\item Meta-interpretation can be a useful technique for building
experimental prototype implementations of knowledge representation
formalisms. In particular, we have shown this for some preferences
formalisms extending the seminal answer set semantics.

\item By the use of answer set programming, it is possible to cast
definitions of the semantics of KR-formalisms quite naturally and
appealingly into extended logic programs, which then, by usage of answer
set programming engines, provide executable specifications.  Note
that, in this line, You {\em et al.} \cite{you-etal-99} have
shown how inheritance networks can be compiled to logic programs, and
that, on the other hand, semantics of logic programs may used for
refining the semantics of inheritance networks.

\item Adding optimization constructs to the basic language of extended
logic programming, such as weak constraints in \dlv and the constructs
provided in Smodels \cite{niem-etal-2000a}, is valuable for elegantly
expressing semantics which are defined in terms of optimal values of
cost functions. The semantics of weakly preferred answer sets provides
a striking example; other examples can be found e.g.\ in the domain of
diagnostic reasoning. Enhancing ASP engines by further constructs and
their efficient realization is thus important for increasing the
usability of the ASP compilation and meta-interpreter approach. 
 
\end{itemize}

Furthermore, the techniques and methods that we used in the design of
the meta-interpreters, in particular the use of ordering relations,
may prove useful for other researchers when designing ASP
implementations of applications.

In conclusion, it appears that meta-interpretation, which is
well-established in Prolog-style logic programming, is also a topic of
interest in ASP, whose exploration also provides useful results for
core ASP itself. We are confident that future work will provide
further evidence for this view.

\subsubsection*{Acknowledgments.}

The authors would like to thank our colleagues and the participants of
the AAAI 2001 Spring Symposium on Answer Set Programming for their
comments on this work. This work was supported by the Austrian Science
Fund (FWF) under grants P13871-INF, %
P14781-INF, and Z29-INF.

{
\newcommand{\SortNoOp}[1]{}

}

\end{document}